\documentclass[12pt,australian,dutch,english,aps,rpm,reprint,onecolumn]{revtex4-2}
\usepackage{tgpagella}
\usepackage{tgheros}
\usepackage[T1]{fontenc}
\usepackage[latin9]{inputenc}
\usepackage{geometry}
\usepackage{babel}
\usepackage{amsmath}
\usepackage{amsthm}
\usepackage{amssymb}
\usepackage{stmaryrd}
\usepackage[unicode=true]
 {hyperref}

\makeatletter

\newcommand{\noun}[1]{\textsc{#1}}

\numberwithin{equation}{section}
\numberwithin{figure}{section}
\theoremstyle{plain}
\newtheorem{thm}{\protect\theoremname}
\theoremstyle{definition}
\newtheorem{defn}[thm]{\protect\definitionname}
\theoremstyle{remark}
\newtheorem*{rem*}{\protect\remarkname}
\theoremstyle{plain}
\newtheorem{cor}[thm]{\protect\corollaryname}
\theoremstyle{plain}
\newtheorem{lem}[thm]{\protect\lemmaname}

\makeatother

\addto\captionsaustralian{\renewcommand{\corollaryname}{Corollary}}
\addto\captionsaustralian{\renewcommand{\definitionname}{Definition}}
\addto\captionsaustralian{\renewcommand{\lemmaname}{Lemma}}
\addto\captionsaustralian{\renewcommand{\remarkname}{Remark}}
\addto\captionsaustralian{\renewcommand{\theoremname}{Theorem}}
\addto\captionsdutch{\renewcommand{\corollaryname}{Corollarium}}
\addto\captionsdutch{\renewcommand{\definitionname}{Definitie}}
\addto\captionsdutch{\renewcommand{\lemmaname}{Lemma}}
\addto\captionsdutch{\renewcommand{\remarkname}{Opmerking}}
\addto\captionsdutch{\renewcommand{\theoremname}{Stelling}}
\addto\captionsenglish{\renewcommand{\corollaryname}{Corollary}}
\addto\captionsenglish{\renewcommand{\definitionname}{Definition}}
\addto\captionsenglish{\renewcommand{\lemmaname}{Lemma}}
\addto\captionsenglish{\renewcommand{\remarkname}{Remark}}
\addto\captionsenglish{\renewcommand{\theoremname}{Theorem}}
\providecommand{\corollaryname}{Corollary}
\providecommand{\definitionname}{Definition}
\providecommand{\lemmaname}{Lemma}
\providecommand{\remarkname}{Remark}
\providecommand{\theoremname}{Theorem}

\begin{document}
\title{A formal theory of experimentation}
\author{Eric Tesse}
\begin{abstract}
A formal theory of experimentation will be presented. Such a theory
ought to be of value, as it presents the necessary \& sufficient conditions
a world must satisfy in order to admit the use of the scientific method.
\end{abstract}
\maketitle

\section{Introduction}

Here, a formal theory of experiments will be developed. If it is successful,
then any scientific theory that is sufficiently expansive to able
to describe experiments ought to conform to this theory. 

In the following section, a general theory of dynamic systems will
be presented. A ``dynamic system'' is any system that can change
with time. This definition will be formalized, elementary operations
will be developed, and a rudimentary concept of knowledge based on
the properties of dynamic systems will be presented. This last will
be particularly significant, because it enables the theory to address
the information gathered by experiments. (Minor note: The term ``dynamic
system'' will be used, rather than the more common ``dynamic\emph{al}
system'', because dynamical systems make assumptions beyond those
that will be required here; specifically, unlike dynamical systems,
the state space will not be required to be a manifold.)

In Sec. \ref{part:Experiments}, dynamic systems will be utilized
to formalize scientific experiments. They will be used to describe
both experimental set-ups as a whole, and the systems whose natures
experiments probe. The analysis will be analogous to that of automata
theory: A sequence of system states is read into the experimental
equipment, and as the states are read in, the equipment's state changes,
until it ultimately enters a final state. The final state determines
which outcome the sequence of states belongs to. In developing this
formalization, we will strive to limit our assumptions to those properties
of experiments that are required by the scientific method.

Finally, in Sec. \ref{part:Probabilities}, a theory of experimental
probability will be presented. Traditional rules of probability and
statistics will be shown to hold for individual experiments. Probability
rules will then be derived for any collection of experiments that
satisfies the constraint that, if any two experiments in the collection
share outcomes, then they agree on those outcomes' probabilities.
The resulting description will be applicable to both classical and
quantum probabilities.

\section{\label{part:Dynamics}Dynamics}

\subsection{Parameters}

In this Section, system dynamics will be formalized in the most general
terms possible. Dynamic systems are systems that can change with time
meaning that, in any mathematical description, they will be \emph{parameterized}
by time. We therefore start by formally defining parameters. Assertions
that follow from the definition will not be explicitly proved, as
the material is likely too familiar to require them. This section
may be skimmed.

Parameters are structures of the form $(\Lambda,<,+,0)$ that satisfy
the following axioms:

Total Ordering:

1) For all $\lambda\in\Lambda$, $\lambda\nless\lambda$

2) For all $\lambda_{1},\lambda_{2},\lambda_{3}\in\Lambda$, if $\lambda_{1}<\lambda_{2}$
and $\lambda_{2}<\lambda_{3}$ then $\lambda_{1}<\lambda_{3}$

3) For all $\lambda_{1},\lambda_{2}\in\Lambda$, either $\lambda_{1}=\lambda_{2}$
or $\lambda_{1}<\lambda_{2}$ or $\lambda_{2}<\lambda_{1}$ 

Additivity:

4) For all $\lambda\in\Lambda$, $\lambda+0=\lambda$

5) For all $\lambda_{1},\lambda_{2}\in\Lambda$, $\lambda_{1}+\lambda_{2}=\lambda_{2}+\lambda_{1}$

6) For all $\lambda_{1},\lambda_{2},\lambda_{3}\in\Lambda$, $(\lambda_{1}+\lambda_{2})+\lambda_{3}=\lambda_{1}+(\lambda_{2}+\lambda_{3})$

Interrelationship between ordering and addition:

7) For all $\lambda_{1},\lambda_{2},\lambda_{3}\in\Lambda$, $\lambda_{1}<\lambda_{2}$
\foreignlanguage{australian}{iff} $\lambda_{1}+\lambda_{3}<\lambda_{2}+\lambda_{3}$ 

8) For all $\lambda_{1},\lambda_{2}\in\Lambda$, if $\lambda_{1}<\lambda_{2}$
then there's a $\lambda_{3}\in\Lambda$ s.t. $\lambda_{1}+\lambda_{3}=\lambda_{2}$ 

Parameters possess a positive element:

9) There exists a $\lambda\in\Lambda$ s.t $\lambda>0$

Multiplication is not defined on parameters. This is because parameters
generally have scale (e.g., values of ``13 sec'', rather than simply
``13'').

A number of basic properties follow immediately from these axioms.
For example, it follows from axioms 7 \& 8 that parameters are either
discrete or dense; they can't be discrete in some places, and dense
in others. It follows from axioms 7, 9, and 4 that parameters are
unbounded from above, and from 7 \& 4 that if a parameter is bounded
from below, its lower bound is $0$. It can also be shown that if
a parameter is unbounded from below, then every parameter value has
an additive inverse. 

Parameters may therefore be categorized into four broad classes, based
on whether they are discrete or dense, and whether or not they are
bounded from below. As things stand, these four classes contain many
structures that do not suit our purposes. For example, the set of
Real numbers whose decimal representation is of finite length satisfies
the above axioms, as does any uncountable limit ordinal. Neither will
be useful for \foreignlanguage{dutch}{parametrizing} dynamic systems.
To eliminate unwanted parameters, we would like to limit our attention
to parameters that satisfy two additional requirement. First, we are
only interested in parameters whose values are finite; this equivalent
to assuming that for any $\lambda_{2}>\lambda_{1}>0$, there exists
an $n\in\mathbb{N}^{+}$ s.t. $\lambda_{1}$ added to itself $n$
times is greater than $\lambda_{2}$. Second, among the dense parameters,
we are only interested in those for which all bounded Cauchy sequences
converge. We therefore add the following two axioms:

10) All subsets of $\Lambda$ that are bounded from above have a least
upper bound. 

11) For any $\lambda_{2}>\lambda_{1}>0$, there exists an $n\in\mathbb{N}^{+}$
s.t. $\lambda_{1}$ added to itself $n$ times is greater than $\lambda_{2}$.

These statements are quite strong; each of the four classes of parameters
now contain only a single element (up to an isomorphism). Those four
parameters are closely related to the four canonical number system,
which can be seen by selecting a parameter value to represent ``$\mathbf{1}$'',
and introducing the operation of multiplication by a number.

First, for each type of parameter, we assign a value to $\mathbf{1}$:
If the parameter is discrete, we assign ``$\mathbf{1}$'' to be
the successor to $0$; if it's dense, we can choose ``$\mathbf{1}$''
to be any parameter value greater than $0$. The choice of $\mathbf{1}$
sets the scale. 

Now to define multiplication by a number. For any $\lambda\in\Lambda$,
define $0\lambda=0$, and for any $n\in\mathbb{N}^{+}$ and any $\lambda\in\Lambda$,
define $n\lambda$ to be the value of $\lambda$ added to itself $n$
times. (With this, axiom 11 can be written ``For any $\lambda_{2}>\lambda_{1}>0$,
there exists an $n\in\mathbb{N}^{+}$ s.t. $n\lambda_{1}>\lambda_{2}$''.)
If the parameter is unbounded from below, then for any $i\in\mathbb{Z}$,
$i<0$, define $i\lambda$ to be the additive inverse of $\bigl|i\bigr|\lambda$.
If the parameter is dense and unbounded from below, for any $\frac{n}{m}\in\mathbb{Q}$,
define $\frac{n}{m}\lambda$ to be the $\lambda^{\prime}$ s.t. $m\lambda^{\prime}=n\lambda$;
if the parameter is bounded from below, limit this operation to $\frac{n}{m}\in\mathbb{Q}^{+}$.
Note that, with the inclusion of axiom 10, $\frac{n}{m}\lambda$ must
exist for dense parameters. 

Axiom 10 further entails that all dense parameters form a continuum,
which means that we ought to be able to multiply dense parameter values
by real numbers as well. This is defined as follows: For any $r\in\mathbb{R}\setminus\mathbb{Q}$
(that is, $r$ is an element of $\mathbb{R}$, but is not an element
of $\mathbb{Q}$), $r>0$ and $\lambda>0$, $r\lambda$ is the least
upper bound of the set $\{q\lambda:q\in\mathbb{Q}^{+}\:and\:q<r\}$.
If the parameter is unbounded from below, then for negative values
of $r$ and/or $\lambda$, $r\lambda$ either is, or is additive inverse
of, $|r||\lambda|$.

This completes the definition of multiplication by the various types
of numbers. Given these, the four types of parameters may be constructed
as follows: If $(\Lambda,<,+,0)$ is discrete and bounded from below,
then $\Lambda=\{n\mathbf{1}:n\in\mathbb{N}\}$, if it's discrete and
unbounded from below, $\Lambda=\{i\mathbf{1}:i\in\mathbb{Z}\}$, if
it's dense and unbounded from below $\Lambda=\{r\mathbf{1}:r\in\mathbb{R}\}$,
and if it's dense and bounded from below $\Lambda=\{r\mathbf{1}:r\in\mathbb{R}^{+}\}$.
In all cases $x_{1}\mathbf{1}+x_{2}\mathbf{1}=(x_{1}+x_{2})\mathbf{1}$
and $x_{1}\mathbf{1}>x_{2}\mathbf{1}$ iff $x_{1}>x_{2}$. 

As indicated above, any parameter of the corresponding type must contain
these sets. Axiom 11 ensures that parameters cannot contain more elements
than those in these constructions. Thus, the above 11 axioms yield
precisely the 4 required parameters. 

In what follows, parameters will generally be represented by their
set of values, $\Lambda$, rather than by their full structure, $(\Lambda,<,+,0)$.

\subsection{\label{sec:State-Transition}State Transitions}

Dynamic systems are systems that change with time. If a system changes
with time, something must be changing; we call whatever it is that
can change the system ``state'', and to describe the ways in which
the system state can evolve over time, we introduce ``transition
chains''. If $\Lambda$ is a system's time parameter, and $\mathcal{P}$
its range of states, then a \emph{transition chain}, $T$, is comprised
of elements of $\Lambda\times\mathcal{P}$ s.t. the system can pass
through each element of $T$ in succession. If the system can pass
from state $s_{1}$ at time $\lambda_{1}$, to state $s_{2}$at $\lambda_{2}$,
and then on to $s_{3}$ at $\lambda_{3}$, then $\{(\lambda_{1},s_{1}),(\lambda_{2},s_{2}),(\lambda_{3},s_{3})\}$
is a transition chain for the system. Transition chains are subject
to the restriction that if $T$ is a transition chain, $(\lambda_{1},s_{1}),(\lambda_{2},s_{2})\in T$,
and $(\lambda_{1},s_{1})\neq(\lambda_{2},s_{2})$, then $\lambda_{1}\neq\lambda_{2}$,
meaning that a chain cannot contain more than one element at any one
time. If a system can undergo multiple transitions simultaneously,
then the dynamics are described by sets of transition chains. Note
that if $T$ is a transition chain, and $T^{\prime}\subset T$, then
$T^{\prime}$ is also a transition chain. 

Let's say that $T=\{(\lambda_{1},s_{1}),(\lambda_{2},s_{2}),(\lambda_{3},s_{3})\}$
is a transition chain, and that there does not exist any chain, $T^{\prime}$,
s.t. $T\subsetneq T^{\prime}$. This must mean that at all other times,
there are no states that the system can transition to, and so for
this transition chain, the system only exists at times $\lambda_{1}$,
$\lambda_{2}$, and $\lambda_{3}$. In such cases, we can enhance
our set of system states to include a state for non-existence, say
$\eta$; if the system is in state $\eta$ at time $\lambda$, it
means that the system doesn't exists at $\lambda$. With the inclusion
this state, then in the above example, there will exist transition
chains, $T^{\prime}$, s.t. $T\subsetneq T^{\prime}$; in the largest
such chain, for all $\lambda$ not in $T$, $(\lambda,\eta)\in T^{\prime}$.
Thus, if the set of states is made sufficiently inclusive, every transition
chain will be a subset of some ``maximal'' chain that possesses
an element at each time. 

Define a system \emph{path} as a total function from $\Lambda$ onto
$\mathcal{P}$ (the range of possible states) whose graph is a transition
chain. A path's graph is a maximal transition chain, and a maximal
transition chain is the graph of a path. Specifying a system's transition
chains is therefore equivalent to specifying its set of possible paths.
For this reason, we will describe dynamic systems in terms of paths.

\subsection{Dynamic Sets \& Spaces}

We can now formally define dynamic systems. 
\begin{defn}
A \emph{dynamic set}, $S$, is a non-empty set of functions s.t. for
some parameter, $\Lambda$, all elements of $S$ have $\Lambda$ as
their domain.

Elements of dynamic sets are referred to as a \emph{path}s. They will
be distinguished by an over bar: $\bar{p}$.

If $S$ is a dynamic set:

For any $\bar{p}\in S$, $\Lambda_{S}\equiv Dom(\bar{p})$. That is,
$\Lambda_{S}$ is the parameter that all elements of $S$ share.

$\mathcal{P}_{S}\equiv\bigcup_{\bar{p}\in S}Ran(\bar{p})$; $\mathcal{P}_{S}$
is the set of \emph{states}.

For $\lambda\in\Lambda_{S}$, $S(\lambda)\equiv\{p\in\mathcal{P}_{S}:for\:some\:\bar{p}\in S,\:\bar{p}(\lambda)=p\}$.
$S(\lambda)$ is the set of possible states at time $\lambda$.

$Uni[S]\equiv\{(\lambda,p)\in\Lambda_{S}\otimes\mathcal{P}_{S}:p\in S(\lambda)\}$;
$Uni[S]$ is the ``universe'' of $S$, the set of all allowed time-state
pairs.
\end{defn}

Dynamic sets encapsulate our basic concepts of system dynamics. To
draw out some of their properties we define two rudimentary operations,
the first one slices paths into pieces, while the second pastes the
pieces together.
\begin{rem*}
We will use the common representation for parameter intervals, with
rounded parenthesis used for open end points, and square brackets
for closed end points. Thus, $(\lambda_{1},\lambda_{2})$ is the set
of all $\lambda$ s.t. $\lambda_{1}<\lambda<\lambda_{2}$ , $(\lambda_{1},\lambda_{2}]$
is the set of all $\lambda$ s.t. $\lambda_{1}<\lambda\leq\lambda_{2}$,
etc. For unbounded intervals, ``$\infty$'' is used in place of
an endpoint; for example, $[-\infty,\lambda_{1}]$ is the set of all
$\lambda$ s.t. $\lambda\leq\lambda_{1}$. If the parameter is bounded
from below, $[-\infty,\lambda]=[0,\lambda]$.
\end{rem*}
\begin{defn}
For dynamic set $S$:

If $\bar{p}\in S$, and $[x_{1},x_{2}]$ is a closed interval of $\Lambda_{S}$,
then $\bar{p}[x_{1},x_{2}]$ is $\bar{p}$ restricted to domain $[x_{1},\,x_{2}]$
(values of $x_{1}=-\infty$ and/or $x_{2}=\infty$ are allowed.) 

$S[x_{1},x_{2}]\equiv\{\bar{p}[x_{1},x_{2}]:\bar{p}\in S\}$

If $\bar{p}_{1},\bar{p}_{2}\in S$, $\lambda\in\Lambda_{S}$, and
$\bar{p}_{1}(\lambda)=\bar{p}_{2}(\lambda)$, then $\bar{p}_{1}[x_{1},\lambda]\circ\bar{p}_{2}[\lambda,x_{2}]$
is the function on domain $[x_{1},x_{2}]$ s.t.:

$\bar{p}_{1}[x_{1},\lambda]\circ\bar{p}_{2}[\lambda,x_{2}](\lambda^{\prime})=\begin{cases}
\bar{p}_{1}(\lambda^{\prime}) & if\,\lambda^{\prime}\in[x_{1},\lambda]\\
\bar{p}_{2}(\lambda^{\prime}) & if\,\lambda^{\prime}\in[\lambda,x_{2}]
\end{cases}$

If $A$ and $B$ are sets of partial paths from $S$ then

$A\circ B\equiv\{\bar{p}_{1}[x_{1},\lambda]\circ\bar{p}_{2}[\lambda,x_{2}]:\bar{p}_{1}[x_{1},\lambda]\in A,\:\bar{p}_{2}[\lambda,x_{2}]\in B,\:and\:\bar{p}_{1}(\lambda)=\bar{p}_{2}(\lambda)\}$ 
\end{defn}

We will refer to the ``$\circ$'' operation as \emph{concatenation}.
For any dynamic set, $S$, any $\lambda\in\Lambda_{S}$, $S\subset S[-\infty,\lambda]\circ S[\lambda,\infty]$.
An interesting property that a dynamic set may possess is closure
under concatenation, i.e., for all $\lambda\in\Lambda_{S}$, $S=S[-\infty,\lambda]\circ S[\lambda,\infty]$.
A dynamic set with this property is called a \emph{dynamic space}.
The definition is equivalent to: If $D$ is a dynamic set, it is a
dynamic space if and only if for all $\lambda\in\Lambda_{D}$, all
$\bar{p}_{1},\bar{p}_{2}\in D$ s.t. $\bar{p}_{1}(\lambda)=\bar{p}_{2}(\lambda)$,
$\bar{p}_{1}[-\infty,\lambda]\circ\bar{p}_{2}[\lambda,\infty]\in D$
(as a rule, we will denote a dynamic space with a ``$D$'' rather
than an ``$S$''). The distinguishing characteristic of dynamic
spaces is that the current state always contains sufficient information
to fully determine the collection of possible future paths, meaning
that information from the system's past never further restricts the
system's collection of possible futures. We will follow the convention
that, unless stated otherwise, closed systems will always be assumed
to be dynamic spaces.

Some other significant properties of dynamic sets involve how their
dynamics may change with time. A dynamic set, $S$, is \emph{homogeneous}
if for all $\bar{p}\in S$, all $\lambda_{1},\lambda_{2}\in\Lambda_{S}$
s.t. $(\lambda_{2},\bar{p}(\lambda_{1}))\in Uni[S]$, there's a $\bar{p}^{\prime}\in S$
s.t. for all $\lambda\in\Lambda_{S}$, $\bar{p}^{\prime}(\lambda)=\bar{p}(\lambda+\lambda_{1}-\lambda_{2})$
(if $\lambda_{1}<\lambda_{2}$, and $\Lambda_{S}$ is bounded from
below, then we instead demand that for all $\lambda\in\Lambda_{S}$,
$\bar{p}(\lambda)=\bar{p}^{\prime}(\lambda+\lambda_{2}-\lambda_{1})$).
$\bar{p}^{\prime}$ is a time-shifted copy of $\bar{p}$. Homogeneity
therefore ensures that the system's behavior is the same regardless
of when it enters a given state. This is not quite enough to say that
the dynamic set does not change with time; for that you also need
to assert that the system's states may be entered at any time. A system
state, $p\in\mathcal{P}_{S}$, is \emph{homogeneously realized} if
for all $\lambda\in\Lambda_{S}$, $p\in S(\lambda)$. Similarly, $A\subset\mathcal{P}_{S}$
is homogeneously realized if every element of $A$ is homogeneously
realized. Thus, a dynamic set's dynamics does not change with time
if it is homogeneous, and $\mathcal{P}_{S}$ is homogeneously realized.

{[}Side Note: A number of interpretations of quantum mechanics posit
that quantum systems exhibit their unusual behavior because they take
all available paths. The introductory material presented here is equally
applicable to systems that take individual paths, and systems that
simultaneously take all available paths. This will continue to be
the case in all that follows.{]}

\subsection{\label{sec:DS Notation}Dynamic Set Notation}

The following simple notation for describing dynamic sets and their
related constructs will prove quite useful.

\smallskip{}

\noun{1. Single condition path expressions}

If $S$ is a dynamic set, and $X\subset\mathcal{P}_{S}$, $S_{\rightarrow X\rightarrow}$
is the set of all elements of $S$ that pass through $X$:

$S_{\rightarrow X\rightarrow}\equiv\{\bar{p}\in S:for\:some\:\lambda\in\Lambda_{S},\:\bar{p}(\lambda)\in X\}$.

\smallskip{}

\noun{2. Adding conditions}

Further sets of states may always be added to a chain of events. Thus,
for example, $S_{\rightarrow X\rightarrow Y\rightarrow Z\rightarrow}$
is the set of paths that pass through $X$, then $Z$, and then $Y$.
Explicitly:

$S_{\rightarrow X\rightarrow Y\rightarrow Z\rightarrow}\equiv\{\bar{p}\in S:for\:some\:\lambda_{1}\leq\lambda_{2}\leq\lambda_{3},\:\bar{p}(\lambda_{1})\in X,\:\bar{p}(\lambda_{2})\in Y,\:and\:\bar{p}(\lambda_{3})\in Z\}$

\smallskip{}

\emph{\noun{3. End Arrows}}

We can switch between sets of paths, semi-paths, and path segments
by adding and removing end arrows from the subscript. For example,
if $X$ and $Y$ are sets of states, $S_{\rightarrow X\rightarrow Y}$
is the set of semi-paths that pass through $X$ and end at $Y$, while
$S_{X\rightarrow Y}$ is the related set of path segments that run
from $X$ to $Y$. Explicitly:

$S_{X\rightarrow Y}\equiv\{\bar{p}[\lambda_{1},\lambda_{2}]:\bar{p}\in S,\:\bar{p}(\lambda_{1})\in X,\:and\:\bar{p}(\lambda_{2})\in Y\}$

$S_{\rightarrow X\rightarrow Y}\equiv\{\bar{p}[-\infty,\lambda_{2}]:\bar{p}\in S\:and\:for\:some\:\lambda_{1}\leq\lambda_{2},\:\bar{p}[\lambda_{1},\lambda_{2}]\in S_{X\rightarrow Y}\}$

$S_{X\rightarrow Y\rightarrow}\equiv\{\bar{p}[\lambda_{1},\infty]:\bar{p}\in S\:and\:for\:some\:\lambda_{2}\geq\lambda_{1},\:\bar{p}[\lambda_{1},\lambda_{2}]\in S_{X\rightarrow Y}\}$

\smallskip{}

\noun{4. Adding Time}

To any set of states, time may be added, indicating paths that pass
through the set of states at the given time. For example $S_{X\rightarrow(\lambda,Y)\rightarrow Z}$
is the set of partial-paths that start at $X$, pass through $Y$
at time $\lambda$, and terminate at $Z$:

$S_{X\rightarrow(\lambda,Y)\rightarrow Z}\equiv\{\bar{p}[\lambda_{0},\lambda_{1}]\in S_{X\rightarrow Z}:\lambda_{0}\leq\lambda\leq\lambda_{1}\:and\:\bar{p}[\lambda_{0},\lambda_{1}](\lambda)\in Y\}$

\smallskip{}

\noun{5. Extended Time Conditions}

Let's say we want to specify that between the time a path passes through
$A$, and the time it passes through $B$, it is always in the set
of states $Z$. We do so by using the following notation:

$S_{A\rightarrow|Z|\rightarrow B}=\{\bar{p}[\lambda_{1},\lambda_{2}]\in S_{A\rightarrow B}:for\:all\:\lambda\in(\lambda_{1},\lambda_{2}),\:\bar{p}(\lambda)\in Z\}$

For end arrows:

$S_{A\rightarrow|Z|\rightarrow}=\{\bar{p}[\lambda_{1}.\infty]\in S_{A\rightarrow}:for\:all\:\lambda>\lambda_{1},\:\bar{p}(\lambda)\in Z\}$

and similarly for $S_{\rightarrow|Z|\rightarrow A}$.

Thus any $\rightarrow$ can be replaced with $\rightarrow|Z|\rightarrow$,
the bracketed set specifying the range of states available for the
trip.

The remaining rules help to simplify the notation under various circumstances.

\smallskip{}

\noun{6. Extended Time Conditions, Pt 2}

In the expression $S_{A\rightarrow|Z|\rightarrow B}$, $|Z|$ applies
to open interval between $A$ and $B$. There may, however, be times
when we want it to apply to one or both of the end points. To specify
this, we replace ``$|$'' with ``$\Vert$''. 

For example: $S_{A\rightarrow|Z\Vert\rightarrow B}\equiv\{\bar{p}[\lambda_{1},\lambda_{2}]\in S_{A\rightarrow B}:for\:all\:\lambda\in(\lambda_{1},\lambda_{2}],\:\bar{p}(\lambda)\in Z\}$
(compare with the definition of $S_{A\rightarrow|Z|\rightarrow B}$).
This may be written in terms of prior notation as $S_{\rightarrow A\rightarrow|Z\Vert\rightarrow B\rightarrow}=S_{\rightarrow A\bigcap B\rightarrow}\bigcup S_{\rightarrow A\rightarrow|Z|\rightarrow B\bigcap Z\rightarrow}$,
where $S_{\rightarrow A\bigcap B\rightarrow}$ contains the paths
in which $A$ and $B$ occur simultaneously (and so the $|Z\Vert$
condition doesn't apply), and $S_{\rightarrow A\rightarrow|Z|\rightarrow B\bigcap Z\rightarrow}$
contributes the paths in which $A$ occurs before $B$.

Similarly, $S_{\rightarrow A\rightarrow\Vert Z|\rightarrow B\rightarrow}\equiv\{\bar{p}[\lambda_{1},\lambda_{2}]\in S_{\rightarrow A\rightarrow B\rightarrow}:for\:all\:\lambda\in[\lambda_{1},\lambda_{2}),\:\bar{p}(\lambda)\in Z\}$
is equal to $S_{\rightarrow A\bigcap B\rightarrow}\bigcup S_{\rightarrow A\bigcap Z\rightarrow|Z|\rightarrow B\rightarrow}$. 

Finally, $S_{\rightarrow A\rightarrow\Vert Z\Vert\rightarrow B\rightarrow}\equiv\{\bar{p}[\lambda_{1},\lambda_{2}]\in S_{\rightarrow A\rightarrow B\rightarrow}:for\:all\:\lambda\in[\lambda_{1},\lambda_{2}],\:\bar{p}(\lambda)\in Z\}$
is simply $S_{\rightarrow A\bigcap Z\rightarrow|Z|\rightarrow B\bigcap Z\rightarrow}$. 

Expressions of this form will come in handy in when discussing experiments.

\smallskip{}

\noun{7. States}

When a set of states contains only a single state, $X=\{p\}$, we
may use $p$ rather than $\{p\}$. For example, rather than $S_{X\rightarrow(\lambda,\{p\})}$,
we may write $S_{X\rightarrow(\lambda,p)}$

\smallskip{}

\noun{8. Promoting the Subscript}

When the dynamic set is understood, or unimportant, we may simply
denote the subscript. For example, we may write $X\!\rightarrow\!(\lambda,Y)\!\rightarrow\!p$
in place of $S_{X\rightarrow(\lambda,Y)\rightarrow p}$. 

\smallskip{}

\noun{9. Negation}

Sometimes we wish to specify what does not happen, rather than what
does happen. This is fairly straightforward; to specify that $Z$,
does not occur, simply specify that $\mathcal{P}_{S}\setminus Z$
does occur. For example, to say that path $\bar{p}$ doesn't pass
through $X$ simply say $\bar{p}\in\:\rightarrow\!|\mathcal{P}_{S}\setminus X|\!\rightarrow$.
To make this easier to read, for any $X\subset\mathcal{P}_{S}$ define
$\widetilde{X}\equiv\mathcal{P}_{S}\setminus X$. The prior expression
then becomes $\bar{p}\in\:\rightarrow\!|\widetilde{X}|\!\rightarrow$.

\subsection{\label{sec:Knowledge}Knowledge}

The above notation can be used to give a straightforward answer to
a question of some significance: given that a dynamic system's current
state is in some set of states, $X$, what is known about the system's
past, and what is known about its future? In this case, $S_{\rightarrow X}$
is the most that can be known the system's past, and $S_{X\rightarrow}$
is the most that can be known about its future, where $S$ is the
set of paths for the dynamic system. In particular, if $s$ is the
state of some component of that system, and $X$ is the set of system
states where the component state is $s$, then $S_{\rightarrow X}$
is what can be known about the past based on the component, and $S_{X\rightarrow}$
is what can be known of the future. This may be anthropomorphized
as: $S_{\rightarrow X}$ is what the component ``knows'' about the
past, and $S_{X\rightarrow}$ is what it knows of the future. This
usage is in keeping with the common assumption that all ``knowledge''
a system possesses at a given time is encoded in its state. For example,
it is generally assumed that everything a person knows is encoded
in their physical brain state. It is certainly the case for computers:
everything in a computer's memory is encoded in the states of its
various memory devices. 

When we speak of things we know to have happened, we often say things
like ``I know I took my keys this morning''. If the system's current
state is in $X$, and $S_{\rightarrow X}=S_{\rightarrow Y\rightarrow X}$,
then it is known that $Y$ happened. Thus, for example, if $X$ is
the set of all world states that contain your current brain state
(or those portions of your brain state utilized for memory), you'd
be justified is claiming that you know $Y$ happened if $S_{\rightarrow X}=S_{\rightarrow Y\rightarrow X}$.
Of course, there is a conceivable scenario in which I clearly remember
taking my keys this morning, and my brain is in a state it would be
in if I had taken my keys, but there is a world path that passes through
a state containing this brain state along which I didn't take my keys.
In this case $S_{\rightarrow X}\neq S_{\rightarrow Y\rightarrow X}$
(where $X$ are world states that contain my current brain state,
and $Y$ are the states in which I take my keys), and my subjective
sense of certain knowledge is not justified. Thus, while we expect
there to be a correlation between our subjective sense of knowledge
and knowledge as defined here, we do not demand that the two must
be identical. In so far as they are different, our subjective knowledge
may be said to lack certainty.

The notation described above can be further utilized. For example,
if the system's current state is in $X$, and $S_{\rightarrow X}=S_{\rightarrow(\lambda,Y)\rightarrow X}$,
then it is known that $Y$ happened at time $\lambda$.

While these concepts are quite rudimentary, they will also prove to
be of considerable value. In particular, assertions like $S_{\rightarrow X}=S_{\rightarrow Y\rightarrow X}$
will be central to our formalization of experiments. Interestingly,
there is little difference between the expression for the knowledge
that $Y$ has happened, $S_{\rightarrow X}=S_{\rightarrow Y\rightarrow X}$,
and the expression for knowing that $Y$ will happen, $S_{X\rightarrow}=S_{X\rightarrow Y\rightarrow}$.
In reality, however, these two types of knowledge are quite different.
We know a great deal more about what has happened than what will happen.
This is not peculiar to us; as we dig into the earth, we uncover information
about prehistoric life and past civilizations, but learn next to nothing
about future life and future civilizations. In a sense, like us, the
earth also remembers its past, but does not know its future. Perhaps
the most elemental example of this asymmetry occurs when we peer into
the night time sky, and can see stars as they existed in the past,
but not as they will be in the future.

Why that may be the case will not concern us here. Experiments, by
their nature, must reliably record events that have occurred. We will
therefore be interested in what we know of the past, given our current
state. There are other activities, such as setting an alarm clock,
that reliably cause events to occur in the future; these are less
common, and will be of less interest to us here.

Finally, a word about closure under concatenation. Because we take
all knowledge to be encoded in the state there is no way, within a
closed system, to tell which $\bar{p}[-\infty,\lambda]\in S_{\rightarrow p}$
the closed system has been following ($p$ being the current state
if the closed system). There is therefore no way to tell, within a
closed system, whether the system has switched which path it is following
due to a concatenation. This is why we are free to follow the convention
that closed system dynamics are closed under concatenation.
\begin{rem*}
It was previously asserted that the formalism being developed will
be equally applicable to systems that take individual paths, and systems
that simultaneously take all available paths. In the current section,
knowledge has been defined in a way that may appear to require the
system to be in a particular state, or in a proper subset of all available
states. To see how the definition may be reconciled with the prior
claim, it only needs to be added that the world is experienced as
a succession of individual states, rather than a multiplicity of simultaneous
states. In the definition of knowledge, the states used in the expressions
may be understood to be the states that you have experienced (or are
experiencing), regardless of whether a system takes an individual
path, or all available paths.
\end{rem*}

\section{\label{part:Experiments}Experiments}

In this part, experiments will be formalized. The formalization will
be somewhat analogous to automata theory. There, a sequence of characters
is read into an automata, which then determines whether the sequence
corresponds to a sentence in a given language. One of the goals of
automata theory is to determine what types languages the various kinds
of automata are capable of deciding. Experiments are similar: System
paths are ``read into'' an experimental set-up, which then determines
which outcome each path belongs to. As in automata theory, one of
our goals will be to determine the sets of outcomes that experiments
are capable of deciding. 

\subsection{Shells}

In this first section, we consider the closed dynamic space that encompasses
both the system being experimented on and all experimental apparatus
(including the experimenter). We use this dynamic space to enforce
some of the basic dynamic properties of experiments, namely, that
experiments are re-runnable, have a clearly defined start, once started
they must complete, and once complete, they ``know'' that the experiment
occurred.

To formalize these requirements we can define experimental \emph{shells}
as triples, $(D,I,F)$, where $D$ is a dynamic space, $I\subset\mathcal{P}_{D}$
is the set of initial states, and $F\subset\mathcal{P}_{D}$ is the
set of final states. We can start by assuming that these satisfy:

1) $D$ is homogeneous

2) $I$ is homogeneously realized

3) $I\!\rightarrow\,=I\!\rightarrow\!F\!\rightarrow$ 

4) $\rightarrow\!F=\:\rightarrow\!I\!\rightarrow\!F$ 

The first two of the above statements ensure that the experiment is
reproducible, the third ensures that when an experiments starts, it
must complete, and the fourth ensures that when the experiment completes,
you know that an experiment took place. This appears to be sufficient
for our stated needs. However, as currently phrased, there is some
ambiguity as to when the experiment starts, and what it ``knows''
once it completes. 

As things stand, a shell can enter initial states multiple times before
it enters a final state, and can enter a succession of final states
at the experiment's completion. This raises a question as to when
the experiment actually starts, and when it ends. For the latter,
we will make the obvious assumption that an experiment ends as soon
as it enters a final state. A path may continue to be in final states
after it first encounters a final state, but once the first encounter
is made, the experimental run has completed.

The situation is less clear for initial states. If an experiment enters
initial states multiple times before entering a final state, when
does the experiment start? Let's begin by assuming that the experiment
can restart without first entering a final state; this is equivalent
to assuming that the start need not correspond to the first time the
shell enters an initial state. Such a restart has the effect of throwing
away information - you ignore everything that happen between the first
initial state, and the actual start. We may assume, without loss of
generality, that the only reason to do this is if the initial data
is, in some sense, bad; for example, the measuring equipment was not
fully warmed up or operational, or the system being experimented on
was not in a desired initial state. Any good data may always be considered
as part of the experimental outcome, including any data that we may
choose to ignore in our later analysis of the outcome. We therefore
assume that any restart is because, at an earlier start, the data
collection was flawed. But if we are not yet collecting good data,
then it's fair to say that we have not yet entered into an experimental
start state; everything being operational ought to be a prerequisite
for being an experimental start state. We therefore reject our earlier
assumption, and assume that the experiment starts the first time it
enters an initial state.

We have not yet ruled out the possibility that the shell may enter
an initial state after the experiment starts, and before it enters
a final state; we have only said that this ought not correspond to
the experiment restarting. However, allowing the experiment to enter
initial states after it starts is awkward. This would be akin to allowing
the experiment to enter a final state prior to the experiment actually
ending (e.g., taking the end of the experiment to be the last time
the experiment enters a final state, rather than the first time).
To avoid confusion, we assume that an experiment only enters an initial
state when it starts, and doesn't enter a final state until it ends.
In practice, this is not difficult to achieve. When the conditions
of an experimental start is reached, the initial state can be tagged
as such; for example, a clock timing the experiment may be set to
$0$. This tagging allows us to clearly identify when the experiment
started.

Due to these considerations, we will require that after an experiment
enters an initial state, it cannot re-enter the set of initial state
until after it has entered a final state (any such re-entry corresponding
to a new run of the experiment). This is equivalent to strengthening
statement (3) from $I\!\rightarrow\,=I\!\rightarrow\!F\!\rightarrow$
to $I\!\rightarrow\,=I\!\rightarrow\!|\tilde{I}\Vert\!\rightarrow\!F\!\rightarrow$
(see Sec \ref{sec:DS Notation} for definitions of ``$\tilde{I}$''
and ``$|\tilde{I}\Vert$''). In this expression, ``$|\tilde{I}\Vert$''
is used rather than ``$|\tilde{I}|$'' for cases in which $I\bigcap F\neq\emptyset$.
We take $p\in I\bigcap F$ to represent an experimental run that stops
as soon as it begins, providing a snapshot of the state of the system
at the time state $p$ was entered. This is allowed, but such a state
should not be reachable during a different experimental run. If we
were to assume $I\!\rightarrow\,=I\!\rightarrow\!|\tilde{I}|\!\rightarrow\!F\!\rightarrow$,
rather than $I\!\rightarrow\,=I\!\rightarrow\!|\tilde{I}\Vert\!\rightarrow\!F\!\rightarrow$,
then for $p_{1}\in I\setminus F$, $p_{2}\in I\bigcap F$, $p_{1}\!\rightarrow\!|\tilde{I}|\!\rightarrow\!p_{2}\!\rightarrow$
may be non-empty, in which case $p_{2}$ is the final state of two
experimental runs, one starting at $p_{1}$, and the other at $p_{2}$.
As discussed earlier, $I\!\rightarrow\!|\tilde{I}\Vert\!\rightarrow\!F\rightarrow\:=(I\bigcap F\!\rightarrow)\bigcup(I\!\rightarrow\!|\tilde{I}|\!\rightarrow\!F\setminus I\!\rightarrow)$,
so using ``$|\tilde{I}\Vert$'' eliminates precisely this scenario. 

One final, somewhat subtle, difficulty needs to be dealt with. There
remain scenarios in which a final state cannot be paired with any
particular experimental run. For example, consider a semi-path, $\bar{p}[-\infty,0]\in\,\rightarrow\!F$,
s.t. for all $n\in\mathbb{N}^{+}$, $\bar{p}(\frac{-1}{2n})\in I$
and $\bar{p}(\frac{-1}{2n+1})\in F$; this satisfies all shell axioms,
but the $F$ state at $\lambda=0$ is not the final state of any particular
experiment. This is because between $\bar{p}(0)\in F$ and any earlier
$I$, there's another $I$, meaning that between any run of the experiment
and $(0,\bar{p}(0))$ there's another experimental run. As we always
want final states to ``know'' about the experimental run that has
just completed, final states must always occur at the end of some
particular experimental run. To ensure this, we strengthen statement
(4) from $\rightarrow\!F=\,\rightarrow\!I\!\rightarrow F$ to $\rightarrow\!F=\,\rightarrow\!I\!\rightarrow\!|\tilde{I}\Vert\!\rightarrow\!F$.
This eliminates the above scenario. As always, the ``$\Vert$''
is for cases where $I\bigcap F\neq\emptyset$.

We now have the complete definition of an experimental shell:
\begin{defn}
A \emph{shell} is a triple, $(D,I,F)$, where $D$ is a dynamic space,
$I,F\subset\mathcal{P}_{D}$ are the sets of initial and final states,
and these satisfy:

1) $D$ is homogeneous

2) $I$ is homogeneously realized

3) $I\!\rightarrow\,=I\!\rightarrow\!|\tilde{I}\Vert\!\rightarrow\!F\!\rightarrow$

4) $\rightarrow\!F=\,\rightarrow\!I\!\rightarrow\!|\tilde{I}\Vert\!\rightarrow\!F$
\end{defn}

With the start and stop properties of experiments in place, we can
turn our attention to what happens during an experiment. To help extract
this information from the shell, we begin by defining the following
set of partial paths:
\begin{defn}
If $S$ is a dynamic set, and $X,Y\subset\mathcal{P}_{S}$ then 

$S_{X/Y}\equiv\{\bar{p}[\lambda_{1},\lambda_{2}]:\bar{p}[\lambda_{1},\infty]\in S_{X\rightarrow Y\rightarrow},\:and\:\lambda_{2}=\inf(\bar{p}^{-1}[\lambda_{1},\infty][Y])\}$
\end{defn}

In the above definition, $\lambda_{2}$ is either the first time $Y$
occurs in $\bar{p}[\lambda_{1},\infty]$ or, if the ``first time''
can't be obtained in the continuum, the moment before $Y$ first appears.
For shell $(D,I,F)$, $D_{I/F}$ is the set of partial-paths representing
what can happen during the course of the experiment. As sets of this
form will be of some importance, the following definition and theorem
will prove useful:
\begin{defn}
\label{def:(X/Y)}If $S$ is a dynamic set and $X,Y\subset\mathcal{P}_{S}$, 

$(X/Y)\equiv\{p\in\mathcal{P}_{S}:for\:some\:\bar{p}[\lambda_{1},\lambda_{2}]\in S_{X/Y},\:p=\bar{p}(\lambda_{2})\}$
\end{defn}

This means that elements in $(X/Y)$ are the end points of the paths
in $S_{X/Y}$.
\begin{thm}
\label{thm:DX/Y reduction}If $D$ is a homogeneous dynamic space
then \textup{$D_{X/Y}=X\!\rightarrow\!\Vert\tilde{Y}|\!\rightarrow\!(X/Y)$}
\end{thm}

\begin{proof}
If $\bar{p}[\lambda_{1},\lambda_{2}]\in D_{X/Y}$ then $\bar{p}(\lambda_{1})\in X$,
$\bar{p}(\lambda_{2})\in(X/Y)$, and for all $\lambda\in[\lambda_{1},\lambda_{2})$,
$\bar{p}(\lambda)\notin Y$, so $D_{X/Y}\subset X\!\rightarrow\!\Vert\tilde{Y}|\!\rightarrow\!(X/Y)$.
(Note that this doesn't require $D$ to be homogeneous, or a dynamic
space.)

It remains to show that $X\!\rightarrow\!\Vert\tilde{Y}|\!\rightarrow\!(X/Y)\subset D_{X/Y}$.
Take $\bar{p}[\lambda_{1},\lambda_{2}]\in X\!\rightarrow\!\Vert\tilde{Y}|\!\rightarrow\!(X/Y)$,
$p\equiv\bar{p}(\lambda_{2})\in(X/Y)$. By the definition of $(X/Y)$,
there's a $\bar{p}^{\prime}[\lambda_{3},\infty]\in p\!\rightarrow$
s.t. $\lambda_{3}=\inf(\bar{p}^{\prime}[\lambda_{3},\infty]^{-1}[Y])$.
By homogeneity, there's a $\bar{p}^{\prime\prime}$ s.t. for all $\lambda\in\Lambda_{D}$,
$\bar{p}^{\prime\prime}(\lambda)=\bar{p}^{\prime}(\lambda+\lambda_{3}-\lambda_{2})$
(assuming $\lambda_{3}\geq\lambda_{2}$ or $\Lambda_{D}$ is unbounded
from below; otherwise $\bar{p}^{\prime}(\lambda)=\bar{p}^{\prime\prime}(\lambda+\lambda_{2}-\lambda_{3})$
may be asserted for all $\lambda\in\Lambda_{D}$). By considering
$\bar{p}[\lambda_{1},\lambda_{2}]\circ\bar{p}^{\prime\prime}[\lambda_{2},\infty]$,
it follows that $\bar{p}[\lambda_{1},\lambda_{2}]\in D_{X/Y}$.
\end{proof}
If $(D,I,F)$ is a shell, $D$ is a homogeneous dynamic space, and
so $I\!\rightarrow\!\Vert\tilde{F}|\!\rightarrow\!(I/F)$ represents
what can happen during the course of the experiment. With this in
mind, we introduce the following sets of experimental states and paths:
\begin{defn}
If $Z=(D,I,F)$ is a shell:

$Int(Z)=\{p\in\mathcal{P}_{D}:I\!\rightarrow\!\Vert\tilde{F}|\!\rightarrow\!p\neq\emptyset\}$
is the\emph{ shell interior}.

$Dom(Z)\equiv F\bigcup Int(Z)$ is the \emph{shell domain}

If $A\subset Dom(Z)$, $\omega_{A}\equiv(0,I)\!\rightarrow\!|\tilde{I}\Vert\!\rightarrow\!A$

If $A\subset F$, $\Theta_{A}\equiv\{\bar{p}[0,\lambda]\in D_{I/F}:for\:some\:\bar{p}^{\prime}[0,\lambda^{\prime}]\in\omega_{A},\:\bar{p}^{\prime}[0,\lambda]=\bar{p}[0,\lambda]\}$
\end{defn}

The first two definitions identify sets of experimental states. The
first set, the shell interior, consists of states that the shell can
enter while the experiment is in progress. In words, $I\!\rightarrow\!\Vert\tilde{F}|\!\rightarrow\!p$
means that $p$ can be reached from $I$ prior to $F$ being reached
(or when $F$ is first reached). From Thm \ref{thm:DX/Y reduction},
together with shell axiom 3, it follows that $Int(Z)$ contains precisely
the states in $D_{I/F}$. The second set, the shell domain, consists
of all states of interest with regard to the experiment.

The second two definitions identify sets of experimental paths. Given
that our current state is in $A$, $\omega_{A}$ tells what has happened
since the experiment started. Because shell dynamics are homogeneous,
and $I$ is homogeneously realized, it is sufficient only consider
paths that start at $\lambda=0$. Finally, given that the shell is
in $A\subset F$, $\Theta_{A}$ captures what occurred during the
experiment just run. It's important to note that this is all that
can be known about what happened. If, for example, we know that the
initial state was in $Y\subsetneq I$, this is because for any $p\in I\setminus Y$,
$p\!\rightarrow\!|\tilde{I}\Vert\!\rightarrow\!A=\emptyset$, and
so this fact is reflected in $\Theta_{A}$. Note too that the experimenter
herself is part of the dynamic space, so if she knows that the experiment
started in $Y$ because she remembers it, and this ought to be reflected
in $A$. 

\subsection{Recorders}

In this section, we split the shell into a system and its environment.
By making several assumptions about the interrelationship between
these two, the shell will take on the character of an experiment in
which the system is observed, and the environment (or parts of the
environment) performs those observations and records the results. 

\subsubsection{Environmental Shells}

Before splitting the shell into a system and its environment, a little
added notation will be helpful. If $X\times Y$ is a Cartesian product,
and $A\subset X\times Y$, the projection of $A$ onto $X$ will be
written as $X\cdot A$, so $X\cdot A=\{x\in X:for\:some\:y\in Y,\:(x,y)\in A\}$;
similarly for projections onto $Y$. For any path, $\bar{p}$, s.t.
$Ran(\bar{p})\subset X\times Y$, $X\cdot\bar{p}$ is the projection
of $\bar{p}$ onto $X$, meaning that for all $\lambda\in\Lambda$,
$(X\cdot\bar{p})(\lambda)=X\cdot(\bar{p}(\lambda))$; similarly for
projections onto $Y$, and for projections of sets of paths onto $X$
and $Y$. ``Environmental shells'' may now be defined.
\begin{defn}
A shell, $Z=(D,I,F)$, is an \emph{environmental shell} if there exist
sets, $\mathcal{E}$ and $\mathcal{S}$, s.t. 

1) $Dom(Z)\subset\mathcal{S}\times\mathcal{E}$

2) $\mathcal{E}\cdot Dom(Z)=\mathcal{E}$ and $\mathcal{S}\cdot Dom(Z)=\mathcal{S}$

3) With $\mathcal{F}\equiv\mathcal{E}\cdot F$, $F=(\mathcal{S}\times\mathcal{F})\bigcap\mathcal{P}_{D}$
\end{defn}

The first assertion states that the shell consists of a system and
its environment, the second that $\mathcal{E}$ \& $\mathcal{S}$
have no excess states, and the third that whether or not you are in
a final state can be determined entirely by the environmental state,
which is in keeping with the assumption that the experimental equipment
resides in the environment.

Strictly speaking, environmental shells are structures with signature
$(D,I,F,\mathcal{S},\mathcal{E})$. However, it is unwieldy to always
have to specify all these sets, so we will fix a standard notation
for environmental shells. For environmental shell $Z$, $\mathbb{D}$
will refer to the shell's dynamic space, $I$ to its initial states,
$F$ to its final states, $\mathcal{S}$ to its system states, $\mathcal{E}$
to its environmental states, and $\mathcal{F}$ will always be $\mathcal{E}\cdot F$.
If we need to specify which shell $I$ (for example) belongs to, we
add a subscript identifying the shell: $I_{Z}$.

Having isolated the system within the shell domain, we may now formally
define experimental outcomes: 
\begin{defn}
If $Z$ is an environmental shell, and $X\subset\mathcal{F}$, $\mathcal{O}_{X}\equiv\mathcal{S}\cdot\Theta_{\mathcal{S}\times X}$
\end{defn}

If an experiment was run, a record was kept of what was measured over
the course of the run, and $X\subset\mathcal{F}$ is the set of environmental
states containing that record, then $\mathcal{O}_{X}\equiv\mathcal{S}\cdot\Theta_{\mathcal{S}\times X}$
is what we know to have occurred in the system during the experiment.

This raises a somewhat subtle point. One of the reasons to run an
experiment is to try to discover the nature of a system's dynamics.
However, a shell contains a dynamic space that includes the system's
dynamics, which appears to imply that we already know what the system
dynamics are. To resolve the apparent conflict, it is best to view
$\mathcal{O}_{\mathcal{F}}$ as containing all system paths considered
possible prior to running the experiment. When we run an experiment,
we have to assume that we understand the workings of the experimental
setup, otherwise we won't know what the state of the experimental
apparatus tells us about what happened in the system being experimented
on. $\mathbb{D}$ only needs to reflect that understanding. In general,
we expect the actual set of allowed paths for the closed system to
be a subset of the shell's dynamic space, $\mathbb{D}$. 

Returning to experimental outcomes, $\mathcal{O}_{X}$ tells us what
happened during the experiment once it has completed. The following
two sets of system paths give us a view into what is happening while
the experiment is in flight:
\begin{defn}
If $A\subset Int(Z)$: 

$\mathcal{\mbox{\ensuremath{\Sigma}}}_{A}\equiv\mathcal{S}\cdot[(0,I)\!\rightarrow\!\Vert\tilde{F}|\!\rightarrow\!A]$

$\Sigma_{A\rightarrow}\equiv\{\bar{s}[\lambda_{1},\lambda_{2}]:for\:some\:\bar{p}[0,\lambda_{2}]\in\Theta_{F},\:\bar{p}(\lambda_{1})\in A\:\&\:\bar{s}[\lambda_{1},\lambda_{2}]=\mathcal{S}\cdot\bar{p}[\lambda_{1},\lambda_{2}]\}$.
\end{defn}

$\Sigma_{A}$ gives the system paths from $I$ to $A$. It is what
you know about what has transpired in the system during the experiment,
given that you know that the current state of $\mathbb{D}$ is in
$A$. $\Sigma_{A\rightarrow}$ gives the system paths from $A$ through
to the end of the experiment.
\begin{thm}
$\Sigma_{(I/F)}=\mathcal{O}_{\mathcal{F}}$ 
\end{thm}

\begin{proof}
$\mathcal{O}_{\mathcal{F}}=\mathcal{S}\cdot\Theta_{F}$. $\Theta_{F}$
is the set of $\bar{p}[0,\lambda]\in D_{I/F}$. Since $D_{I/F}=I\!\rightarrow\!\Vert\tilde{F}|\!\rightarrow\!(I/F)$,
$\Theta_{F}=(0,I)\!\rightarrow\!\Vert\tilde{F}|\!\rightarrow\!(I/F)$.
Therefore $\mathcal{O}_{\mathcal{F}}=\mathcal{S}\cdot[(0,I)\!\rightarrow\!\Vert\tilde{F}|\!\rightarrow\!(I/F)]=\Sigma_{(I/F)}$.
\end{proof}
In order for an environmental shell to behave as a well-constructed
experiment, the interactions between the system and its environment
must respect some constraints. It is to these that we will turn next.

\paragraph*{Note on quantum mixed states:}

There may be concern that environmental shells are too narrow a construction,
because they do not include quantum cases in which the overall state
cannot be written as the product of a system state \& an environmental
state. Such cases are prevalent when modeling measurements within
quantum theory. For example, imagine a particle that can be either
spin up or spin down, and an environmental state, $\left|E\right\rangle $,
that will evolve to $\left|A\right\rangle $ if it interacts with
a spin up particle, and to $\left|B\right\rangle $ if it interacts
with a spin down particle: $\left|E\right\rangle (\alpha\left|\shortuparrow\right\rangle +\beta\left|\shortdownarrow\right\rangle )\rightarrow\alpha\left|A\right\rangle \left|\shortuparrow\right\rangle +\beta\left|B\right\rangle \left|\shortdownarrow\right\rangle $.
The final state cannot be factored into a system state and an environment
state.

None-the-less, when scientists experience the performance an experiment,
the experimental set-up as a whole can always be decomposed into the
system being experimented on and its environment. This separation
is necessary in order to preserve the identity of the system whose
nature is being probed. It is this that we wish to capture with environmental
shells. A scientific theory, if it is complete, should also be able
to describe experiments as they are experienced by scientists; a theory
does not have to directly speak the language of environmental shells,
but a complete theory ought to be able to reproduce it.

In traditional quantum theory, the experimenter's experience of measurement
is often described in terms of decoherence. In the above measurement
example, take the partial trace over system states to eliminate aspects
of the state that the experimenter does not directly experience. That
yields $\left|E\right\rangle \left\langle E\right|\rightarrow\left|\alpha\right|^{2}\left|A\right\rangle \left\langle A\right|+\left|\beta\right|^{2}\left|B\right\rangle \left\langle B\right|$,
where environmental state $\left|A\right\rangle $ is associated with
particle state $\left|\shortuparrow\right\rangle $, and environmental
state $\left|B\right\rangle $ is associated with $\left|\shortdownarrow\right\rangle $.
This can be interpreted as meaning that the experimenter will experience
the set-up as whole as either evolving to state $\left|A\right\rangle \left|\shortuparrow\right\rangle $,
or to state $\left|B\right\rangle \left|\shortdownarrow\right\rangle $,
in agreement with experience (and with environmental shells). 

Note that even if this interpretation of the partial trace is accepted,
other questions about the measurement may remain open. For example,
in order to know that the particle is in state $\left|\shortuparrow\right\rangle $,
it is not enough to know that the final state of the environment is
$\left|A\right\rangle $, it must also be known that its initial state
was $\left|E\right\rangle $. But how is that known? If $\gamma\left|A\right\rangle \left|\shortdownarrow\right\rangle +\delta\left|B\right\rangle \left|\shortuparrow\right\rangle $
is also an allowed state, and that state evolves from initial state
$\left|\Psi\right\rangle =\alpha_{1}\bigl|E_{1}\bigr\rangle\bigl|\shortuparrow\bigr\rangle+\alpha_{2}\bigl|E_{2}\bigr\rangle\bigl|\shortdownarrow\bigr\rangle$,
then knowing that the environment ended up in state $\left|A\right\rangle $
would not alone allow the experimenter to remember that the initial
environmental state was $\left|E\right\rangle $, or to know that
the current system state is $\left|\shortuparrow\right\rangle $.
That does not reflect how scientists experience experiments. If this
apparent discrepancy cannot be rectified, it would represent an explanatory
gap in the theory, not a flaw in experimentalist's descriptions of
their experiments.

In consequence, the aim of the current formalization of experiments
is to formalize experiments as described by experimentalists, particularly
those aspects that appear to be required by the scientific method.

\subsubsection{Unbiased Conditions}

In an experiment, the system's environment ought not influence the
experimental outcome; an environmental shell is \emph{unbiased} if
this is the case. In its strongest form, this demands that different
environmental states cannot effect the system differently. Naively,
this may be written: For any pair of states sharing the same system
state, $(s,e_{1}),(s.e_{2})\in Int(Z)$, $\Sigma_{(s,e_{1})\rightarrow}=\Sigma_{(s,e_{2})\rightarrow}$.
However, because in general $\Sigma_{(s,e_{1})}\neq\Sigma_{(s,e_{2})}$,
the elements of $\Sigma_{(s,e_{1})\rightarrow}$ and $\Sigma_{(s,e_{2})\rightarrow}$
may terminate at different times. A further complication is that,
within $\Theta_{F}$, $(s,e_{1})$ and $(s,e_{2})$ may not be realized
at the same times; because $\mathcal{O}_{\mathcal{F}}$ does not need
to be homogeneous, if $(s,e_{1})$ and $(s,e_{2})$ are realized at
different times, they need not have similar $\Sigma_{(s,e_{i})\rightarrow}$'s.
These considerations lead to a definition whose wording is more convoluted,
but whose meaning is essentially the same. 
\begin{defn}
An environmental shell is \emph{strongly unbiased} if for every $\lambda\in Dom(\Theta_{F})$,
every $(s,e_{1}),(s,e_{2})\in\Theta_{F}(\lambda)$, $\bar{s}_{1}[\lambda,\lambda_{1}]\in\Sigma_{(s,e_{1})\rightarrow}$
iff there exists a $\bar{s}_{2}[\lambda,\lambda_{2}]\in\Sigma_{(s,e_{2})\rightarrow}$
s.t., with $\lambda^{\prime}\equiv Min(\lambda_{1},\lambda_{2})$,
$\bar{s}_{1}[\lambda,\lambda^{\prime}]=\bar{s}_{2}[\lambda,\lambda^{\prime}]$.
\end{defn}

This condition means that any effect the environment may have on the
system can be incorporated into the system dynamics, allowing the
system to be comprehensible without having to explicitly reference
its environment. The following theorem shows that if an environmental
shell is strongly unbiased, $\mathcal{O}_{\mathcal{F}}$ behaves like
a dynamic space. 
\begin{thm}
\label{thm:If unbiased then dynamic space} If an environmental shell
is strongly unbiased, for every $\bar{s}_{1}[0,\lambda_{1}],\bar{s}_{2}[0,\lambda_{2}]\in\mathcal{O}_{\mathcal{F}}$,
if $\bar{s}_{1}(\lambda)=\bar{s}_{2}(\lambda)$ ($\lambda\leq\lambda_{1}$
and $\lambda\leq\lambda_{2}$), then there exists a $\bar{s}_{1}[0,\lambda]\circ\bar{s}_{3}[\lambda,\lambda_{3}]\in\mathcal{O}_{F}$
s.t., with $\lambda_{m}\equiv Min(\lambda_{2},\lambda_{3})$, $\bar{s}_{3}[\lambda,\lambda_{m}]=\bar{s}_{2}[\lambda,\lambda_{m}]$. 
\end{thm}

\begin{proof}
First note that, regardless of whether the environmental shell is
unbiased, for any $p\in Int(Z)$, if $\bar{s}[0,\lambda]\in\Sigma_{p}$
and $\bar{s}^{\prime}[\lambda,\lambda^{\prime}]\in\Sigma_{p\rightarrow}$
then $\bar{s}[0,\lambda]\circ\bar{s}^{\prime}[\lambda,\lambda^{\prime}]\in\mathcal{O}_{\mathcal{F}}$.
This is because $p$ is a state of the closed system, and the closed
system is a dynamic space.

Now let's turn to the case stated in the theorem. With $s=\bar{s}_{1}(\lambda)$,
for some $e_{1},e_{2}\in\mathcal{E}$, $\bar{s}_{1}[0,\lambda]\in\Sigma_{(s,e_{1})}$
and $\bar{s}_{2}[\lambda,\lambda_{2}]\in\Sigma_{(s,e_{2})\rightarrow}$.
Since the shell is unbiased, there is a $\bar{s}_{3}[\lambda,\lambda_{3}]\in\Sigma_{(s,e_{1})\rightarrow}$
s.t. $\bar{s}_{3}[\lambda,\lambda_{m}]=\bar{s}_{2}[\lambda,\lambda_{m}]$
($\lambda_{m}\equiv Min(\lambda_{2},\lambda_{3})$). By the considerations
of the prior paragraph, $\bar{s}_{1}[0,\lambda]\circ\bar{s}_{3}[\lambda,\lambda_{3}]\in\mathcal{O}_{F}$.
\end{proof}
There are experiments for which the strong unbiased condition fails,
but the experiment is still accepted as valid. Consider the case in
which a particle's position is measured at time $\lambda_{1}$, and
if the particle is found to be in region $A$, the particle's spin
will be measured along the y-axis at $\lambda_{2}$, and if particle
is not in region $A$, the spin will be measured along the z-axis
at $\lambda_{2}$. Assume that there are a pair of particle paths,
$\bar{s}_{1}$ and $\bar{s}_{2}$, s.t. $\bar{s}_{1}$ is in region
$A$ at $\lambda_{1}$, $\bar{s}_{2}$ is not in region $A$ at $\lambda_{1}$,
and for some $\lambda\in(\lambda_{1},\lambda_{2})$, $\bar{s}_{1}(\lambda)=\bar{s}_{2}(\lambda)$.
$\bar{s}_{1}[-\infty,\lambda]\circ\bar{s}_{2}[\lambda,\infty]$ is
not a possible path because a particle can't be in region $A$ at
$\lambda_{1}$ and have its spin polarized along the z-axis at $\lambda_{2}$.
Therefore the set of particle paths is not a dynamic space. Since
the strong unbiased condition ensures that the system can be described
by a dynamic space, that condition must have failed.

To see the problem, assume that if the particle takes path $\bar{s}_{1}$,
the environmental shell will be in state $(s,e_{1})$ at $\lambda$,
whereas if it takes path $\bar{s}_{2}$ it will be in state $(s,e_{2})$
at $\lambda$. $e_{1}$ and $e_{2}$ determine different futures for
the particle: $e_{1}$ ensures that the spin will be measured along
the y-axis at $\lambda_{2}$ while $e_{2}$ ensures that the spin
will be measured along the z-axis. This violates the strong unbiased
condition. However, it doesn't necessarily create a problem for the
experiment because $e_{1}$ and $e_{2}$ ``know'' enough about the
particle's past to know $s_{1}[-\infty,\lambda]$ and $s_{2}[-\infty,\lambda]$
must reside in separate outcomes (one belongs to the set of ``in
$A$ at $\lambda_{1}$'' outcomes, the other to the set of ``not
in $A$ at $\lambda_{1}$'' outcomes). Once paths have been differentiated
into separate outcomes, they may be treated differently by the environment.
This motivates a weak version of the unbiased condition, one that
only holds when $(s,e_{1})$ and $(s,e_{2})$ have not yet been differentiated
into separate outcomes.

To formulate this weaker version, we need to understand the conditions
under which two states in $Int(Z)$ are fated to result in different
outcomes. We start by capturing what is known about experimental outcomes
as of some time $\lambda$. This concept will prove to be of broad
value.
\begin{defn}
\label{def: |O=00005B=00005D|}For environmental shell $Z$, $e\in\mathcal{F}$,
$\lambda\geq0$

$|\mathcal{O}_{e}[0,\lambda]|_{0}\equiv\mathcal{O}_{e}[0,\lambda]$

$|\mathcal{O}_{e}[0,\lambda]|_{n+1}\equiv\{\bar{s}[0,\lambda]:for\:some\:e^{\prime}\in\mathcal{F},\:\mathcal{O}_{e^{\prime}}[0,\lambda]\bigcap|\mathcal{O}_{e}[0,\lambda]|_{n}\neq\emptyset,\:\&\:\bar{s}[0,\lambda]\in\mathcal{O}_{e^{\prime}}[0,\lambda]\}$

$|\mathcal{O}_{e}[0,\lambda]|\equiv\bigcup_{n\in\mathbb{N}}|\mathcal{O}_{e}[0,\lambda]|_{n}$
\end{defn}

If $\bar{s}[0,\lambda]\in|\mathcal{O}_{e}[0,\lambda]|$ and $\bar{s}^{\prime}[0,\lambda]\in\mathcal{O}_{\mathcal{F}}\setminus|\mathcal{O}_{e}[0,\lambda]|$,
then $\bar{s}[0,\lambda]$ and $\bar{s}^{\prime}[0,\lambda]$ are
heading to different outcomes. To extend this to the states in $Int(Z)$,
we use the following theorem:
\begin{thm}
\label{thm:pre-undiff}For environmental shell $Z$, given any $e,e^{\prime}\in\mathcal{F}$,
and any $\lambda\in\Lambda_{\mathbb{D}}$ s.t. $\mathcal{O}_{e^{\prime}}[0,\lambda]\bigcap|\mathcal{O}_{e}[0,\lambda]|=\emptyset$,
for all $p\in Int(Z)$, if $\mathcal{\mbox{\ensuremath{\Sigma}}}_{p}\bigcap|\mathcal{O}_{e}[0,\lambda]|\neq\emptyset$
then $\mathcal{\mbox{\ensuremath{\Sigma}}}_{p}\bigcap\mathcal{O}_{e^{\prime}}[0,\lambda]=\emptyset$
\end{thm}

\begin{proof}
Take any $\mathcal{O}_{e^{\prime\prime}}[0,\lambda]$ s.t. $\mathcal{\mbox{\ensuremath{\Sigma}}}_{p}\bigcap\mathcal{O}_{e^{\prime\prime}}[0,\lambda]\neq\emptyset$.
For this to hold, there must then be a $\bar{p}^{\prime}[\lambda,\lambda^{\prime}]\in(\lambda,p)\!\rightarrow\!|\tilde{I}\Vert\!\rightarrow\!e^{\prime\prime}$.
Remembering that $\mathcal{\mbox{\ensuremath{\Sigma}}}_{p}=\mathcal{S}\cdot[(0,I)\!\rightarrow\!\Vert\tilde{F}|\!\rightarrow\!p]$,
and that $\mathbb{D}$ is a dynamic space, it then follows that for
any $\bar{s}[0,\lambda]\in\mathcal{\mbox{\ensuremath{\Sigma}}}_{p}$,
$\bar{s}[0,\lambda]\in\mathcal{O}_{e^{\prime\prime}}[0,\lambda]$.
Because $\mathcal{\mbox{\ensuremath{\Sigma}}}_{p}\bigcap|\mathcal{O}_{e}[0,\lambda]|\neq\emptyset$,
it follows from the definition of $|\mathcal{O}_{e}[0,\lambda]|$
that $\mathcal{O}_{e^{\prime\prime}}[0,\lambda]\subset|\mathcal{O}_{e}[0,\lambda]$.
\end{proof}
Thus, a $\mathcal{\mbox{\ensuremath{\Sigma}}}_{p}$ can only intersect
a single $|\mathcal{O}_{e}[0,\lambda]|$; all paths in $\mathcal{\mbox{\ensuremath{\Sigma}}}_{p}$
that are not elements of $|\mathcal{O}_{e}[0,\lambda]|$ end at times
other than $\lambda$. We can now define when two states are undifferentiated:
\begin{defn}
For environmental shell, $Z$, $p,p^{\prime}\in Int(Z)$ are \emph{undifferentiated
at $\lambda$} if for some $e\in\mathcal{F}$, $\Sigma_{p}\bigcap|\mathcal{O}_{e}[0,\lambda]|\neq\emptyset$
and $\Sigma_{p^{\prime}}\bigcap|\mathcal{O}_{e}[0,\lambda]|\neq\emptyset$.
\end{defn}

By Thm \ref{thm:pre-undiff}, if $p_{1}$ \& $p_{2}$ are undifferentiated
at $\lambda$, and $p_{2}$ \& $p_{3}$ are undifferentiated at $\lambda$,
then $p_{1}$ \& $p_{3}$ are undifferentiated at $\lambda$. The
definition therefore divides the elements of $\Theta_{F}(\lambda)$
into equivalence classes. We can now define the weak unbiased condition
in the same way as the strong condition, except that it only applies
to undifferentiated states.
\begin{defn}
An environmental shell is \emph{weakly unbiased} if for every $\lambda\in Dom(\Theta_{F})$,
every $(s,e_{1}),(s.e_{2})\in\Theta_{F}(\lambda)$ s.t. $(s,e_{1})$
and $(s,e_{2})$ are undifferentiated at $\lambda$, $\bar{s}_{1}[\lambda,\lambda_{1}]\in\Sigma_{(s,e_{1})\rightarrow}$
iff there exists a $\bar{s}_{2}[\lambda,\lambda_{2}]\in\Sigma_{(s,e_{2})\rightarrow}$
s.t., with $\lambda^{\prime}\equiv Min(\lambda_{1},\lambda_{2})$,
$\bar{s}_{1}[\lambda,\lambda^{\prime}]=\bar{s}_{2}[\lambda,\lambda^{\prime}]$.
\end{defn}

If the weakly unbiased condition holds, the system may not behave
like a dynamic space. However, the following weak version of Thm \ref{thm:If unbiased then dynamic space}
does hold.
\begin{thm}
\label{thm:If unbiased then autonomous}For any weakly unbiased environmental
shell, for every $\bar{s}_{1}[0,\lambda_{1}]\in\mathcal{O}_{\mathcal{F}}$,
every $\lambda\leq\lambda_{1}$, if \textup{$\bar{s}_{1}[0,\lambda],\bar{s}_{2}[0,\lambda]\in|\mathcal{O}_{e}[0,\lambda]|$
(for some $e\in\mathcal{F}$), and }$\bar{s}_{1}(\lambda)=\bar{s}_{2}(\lambda)$,\textup{
then there exists a }$\bar{s}_{2}[0,\lambda]\circ\bar{s}[\lambda,\lambda_{2}]\in\mathcal{O}_{\mathcal{F}}$
s.t., with $\lambda_{m}\equiv Min(\lambda_{1},\lambda_{2})$, $\bar{s}[\lambda,\lambda_{m}]=\bar{s}_{1}[\lambda,\lambda_{m}]$.
\end{thm}

\begin{proof}
Choose $e_{1},e_{2}\in\mathcal{E}$ s.t. $\bar{s}_{2}[0,\lambda]\in\Sigma_{(s,e_{2})}$,
$\bar{s}_{1}[\lambda,\lambda_{1}]\in\Sigma_{(s,e_{1})\rightarrow}$,
and $\Sigma_{(s,e_{1})}\bigcap|\mathcal{O}_{e}[0,\lambda]|\neq\emptyset$.
Note that $\Sigma_{(s,e_{2})}\bigcap|\mathcal{O}_{e}[0,\lambda]|\neq\emptyset$.
Such $e_{1}$ and $e_{2}$ must exist because $\bar{s}_{1}[0,\lambda],\bar{s}_{2}[0,\lambda]\in|\mathcal{O}_{e}[0,\lambda]|$.
$(s,e_{1})$ and $(s,e_{2})$ are undifferentiated, and so the proof
now proceeds similarly to that for Thm \ref{thm:If unbiased then dynamic space}.
\end{proof}

\subsubsection{\label{subsec:Decidability}Recorders \& Decidability}

Weakly unbiased environmental shells represent our base formalization
of experiment. Due to their environment's ability to record events
without actively participating in them, we refer to them as ``recorders'':
\begin{defn}
A \emph{recorder} is a weakly unbiased environmental shell\emph{.}
\end{defn}

We now consider what a recorder can tell us about its system. As things
stand, recorders only give us information about what occurs in the
dynamic set shard $\mathcal{O}_{\mathcal{F}}$. As we would like to
represent the system being experimented on by a full dynamic set,
we would like to be able to translate recorder outcomes into information
about dynamic sets. To do so, we start with the following definitions:

For any set of partial paths, $A$, any dynamic set, $S$, define
$S_{\rightarrow A\rightarrow}\equiv\{\bar{p}\in S:for\:some\:\bar{p}^{\prime}[x_{1},x_{2}]\in A,\:\bar{p}[x_{1},x_{2}]=\bar{p}^{\prime}[x_{1},x_{2}]\}$,
and similarly for $S_{A\rightarrow}$ and $S_{\rightarrow A}$.

For any recorder, any $\lambda\in\Lambda_{\mathbb{D}}$, $A\subset F$,
and $X\subset\mathcal{F}$, define $\Theta_{A}^{\lambda}$ and $\mathcal{O}_{X}^{\lambda}$
to be the analogues of $\Theta_{A}$ and $\mathcal{O}_{X}$ for experiments
that start at time $\lambda$ rather than $0$. Formally, this means
$\Theta_{A}^{\lambda}\equiv\{\bar{p}[\lambda,\lambda^{\prime}]\in\mathbb{D}_{I/F}:for\:some\:\bar{p}^{\prime}[\lambda,\lambda^{\prime\prime}]\in(\lambda,I)\!\rightarrow\!|\tilde{I}\Vert\!\rightarrow\!A,\:\bar{p}^{\prime}[\lambda,\lambda^{\prime}]=\bar{p}[\lambda,\lambda^{\prime}]\}$,
and $\mathcal{O}_{X}^{\lambda}\equiv\mathcal{S}\cdot\Theta_{\mathcal{S}\times X}^{\lambda}$.
Because $\mathbb{D}$ is homogeneous and $I$ is homogeneously realized,
$\Theta_{A}^{\lambda}$ and $\mathcal{O}_{X}^{\lambda}$ are simply
$\Theta_{A}$ and $\mathcal{O}_{X}$ shifted by $\lambda$.

Finally, if $X$ is a set, $C$ is a \emph{covering} of $X$ if it
is a collection of nonempty subsets of $X$ and $\bigcup C=X$. (Note:
it is more common to demand that $X\subset\bigcup C$; the restriction
to $\bigcup C=X$ will allow for some mild simplification.) $C$ is
a \emph{partition} of $X$ if it is a pairwise disjoint covering.

With these, we can state what records reveal about a system's dynamic
set:
\begin{defn}
A covering, $K$, of dynamic set $S$ is \emph{decided} by recorder
$Z$ if for some $\lambda\in\Lambda_{S}$, $K=\{S_{\rightarrow\mathcal{O}_{e}^{\lambda}\rightarrow}:e\in\mathcal{F}\:and\:S_{\rightarrow\mathcal{O}_{e}^{\lambda}\rightarrow}\neq\emptyset\}$. 
\end{defn}

We do not demand that every partial path in $\mathcal{O_{F}^{\lambda}}$
must correspond to some path in $S$. This is because, as mentioned
above, we allow $\mathcal{O_{F}^{\lambda}}$ to be overstuffed with
paths. If we wish every element of $\mathcal{O_{F}^{\lambda}}$ to
correspond to a path in $S$, we add the condition ``for every $\bar{s}[\lambda,\lambda^{\prime}]\in\mathcal{O_{F}^{\lambda}}$,
$\bar{s}[\lambda,\lambda^{\prime}]\in S[\lambda,\lambda^{\prime}]$''.

In general, recorders decide coverings rather than partitions. It
is often more natural to think of experimental outcomes as partitioning
the set of experimental paths, rather than covering it. Fortunately,
from any covering, a partition may be derived. To do so, we'll extend
the definition of $|\mathcal{O}_{e}[0,\lambda]|$ given above. 
\begin{defn}
\label{def:|...|}If $K$ is a covering of a dynamic set, and $\alpha\in K$:

$|\alpha[\lambda_{1},\lambda_{2}]|_{0}\equiv\alpha$

$|\alpha[\lambda_{1},\lambda_{2}]|_{n+1}\equiv\{\bar{s}[\lambda_{1},\lambda_{2}]:for\:some\:\beta\in K,\:\beta[\lambda_{1},\lambda_{2}]\bigcap|\alpha[\lambda_{1},\lambda_{2}]|_{n}\neq\emptyset,\:\&\:\bar{s}[\lambda_{1},\lambda_{2}]\in\beta[\lambda_{1},\lambda_{2}]\}$

$|\alpha[\lambda_{1},\lambda_{2}]|\equiv\bigcup_{n\in\mathbb{N}}|\alpha[\lambda_{1},\lambda_{2}]|_{n}$

$|K[\lambda_{1},\lambda_{2}]|=\{|\alpha[\lambda_{1},\lambda_{2}]|:\alpha\in K\}$

Finally, $|\alpha|\equiv|\alpha[-\infty,\infty]|$ and $|K|\equiv|K[-\infty,\infty]|$
\end{defn}

$|\alpha[-\infty,\lambda]|$ and $|K[-\infty,\lambda]|$ will come
in handy shortly, but for now we only need $|\alpha|$ and $|K|$.
$|K|$ is clearly a partition, and is the most fine grained partition
that can be formed from $K$. If $Z$ decides $K$, then for any $x\in|K|$,
every run of $Z$ will determine whether or not $x$ occurred. 

There is an important sense in which the above definition of decidability
can be considered to be too broad. Closed systems are always dynamic
spaces, so if a system is described by a dynamic set that is not a
dynamic space, this is due to interactions with its environment. In
the above definition, interactions between the system and the environment
are regulated via the unbiased condition while the experiment is taking
place, but before and after the experiment, any kind of interactions
are allowed, even those that would undermine the autonomy of the system.
It is more sensible to assume that these interactions are always unbiased.
Indeed, if we make the natural assumption that the system and environment
don't interact at all outside of the experiment, then it follows immediately
that the interactions are always unbiased. 

To formalize this, we note that if nothing is measured prior to $\lambda$,
then prior to $\lambda$ there is no distinction between being weakly
\& strongly unbiased and so Thm \ref{thm:If unbiased then dynamic space}
ought to hold, and from $\lambda$ onward, Thm \ref{thm:If unbiased then autonomous}
ought to hold. This gives us:
\begin{defn}
If $S$ is a dynamic set, $K$ is a covering of $S$, and $\lambda\in\Lambda_{S}$,
\emph{$S$ is unbiased with respect to $(\lambda,K)$} if

1) For all $\lambda^{\prime}\leq\lambda$, $S=S[-\infty,\lambda^{\prime}]\circ S[\lambda^{\prime},\infty]$

2) For all $a\in K$, all $\lambda^{\prime}>\lambda$, all $\bar{p}_{1},\bar{p}_{2}\in S_{|a[-\infty,\lambda^{\prime}]|\rightarrow}$,
if $\bar{p}_{1}(\lambda^{\prime})=\bar{p}_{2}(\lambda^{\prime})$
then $\bar{p}_{1}[-\infty,\lambda^{\prime}]\circ\bar{p}_{2}[\lambda^{\prime},\infty]\in S$.
\end{defn}

This effectively enforces the unbiased condition for all times. With
it, we are in a position to define a more rigorous form of decidability.
\begin{defn}
A covering, $K$, of dynamic set $S$ is \emph{properly decided} by
recorder $Z$ if, for some $\lambda\in\Lambda_{S}$, $S$ is unbiased
with respect to $(\lambda,K)$ and $K=\{S_{\rightarrow\mathcal{O}_{e}^{\lambda}\rightarrow}:e\in\mathcal{F}\:and\:S_{\rightarrow\mathcal{O}_{e}^{\lambda}\rightarrow}\neq\emptyset\}$.
\end{defn}

If $S$ is a dynamic space then $S$ is unbiased with respect to any
\emph{$(\lambda,K)$}, so for dynamic spaces, ``decidable'' and
``properly decidable'' are identical. 

The following holds trivially:
\begin{thm}
\label{thm:trivial}If covering $K$ of dynamic set $S$ is properly
decided by a recorder, then 

1) There's a $\lambda_{0}\in\Lambda_{S}$ s.t. for all $\alpha\in K$,
all $\lambda\leq\lambda_{0}$, $\alpha=S[-\infty,\lambda]\circ\alpha[\lambda,\infty]$

2) For all $\alpha\in K$ there's a $\lambda_{\alpha}\in\Lambda_{S}$
s.t. $\alpha=S_{\alpha[-\infty,\lambda_{\alpha}]\rightarrow}$
\end{thm}

\subsection{Ideal Recorders}

For many purposes, recorders are a weak of a construction. In particular,
recorders do not allow us to draw a clean distinction between two
different types of uncertainty: uncertainty about the system given
complete knowledge of the environment (intrinsic uncertainty), and
uncertainty about the state of the environment, particularly with
regard to knowing precisely which $e\in\mathcal{F}$ occurred (extrinsic
uncertainty). Practically speaking, it is often necessary to treat
these two separately. The matter is particularly pressing in quantum
theory, because there these two types of uncertainty are associated
with two different methods for calculating probability. 

In this section, two properties will be introduced that together will
remove the ambiguity between these types of uncertainty. No further
refinements to recorders that will be required.

\subsubsection{All-Reet Recorders}

It is possible for the environment to record a piece of information
about the system at one time, and then later ``forget'' it, so that
it is not ultimately reflected in the experimental outcome. This would
lead to ambiguity as to how to describe the final outcome. For example,
if at time $\lambda$ the environment records whether the system is
in region $A$, $B$, or $C$, but by the end of the experiment, the
final environmental state can only reveal whether the system was in
region $A\bigcup B$ or $C$, should the first outcome be understood
to be ``$A\bigcup B$'' (intrinsic uncertainty), or ``$A\:or\:B$''
(extrinsic uncertainty), or something distinct from either? 

To eliminate this ambiguity, we introduce recorders that never forget.
Let's start by noting that if $\mathcal{E}\cdot\Theta_{e}$ has only
one element, then the outcome, $\mathcal{O}_{e}$, has retained all
the information ever gained about the system. For example, if $\mathcal{E}\cdot\Theta_{e}=\{\bar{e}[0,\lambda]\}$,
and $\bar{e}[0,\lambda](\lambda^{\prime})=e^{\prime}$, then final
state, $e$, knows that the environment was in state $e^{\prime}$
at time $\lambda^{\prime}$, and so knows everything about the system
that was known by the environment at time $\lambda^{\prime}$. This
rule may be generalized: if $\Theta_{e}=(\mathcal{S}\cdot\Theta_{e})\times(\mathcal{E}\cdot\Theta_{e})=\mathcal{O}_{e}\times(\mathcal{E}\cdot\Theta_{e})$,
then every environmental path contains the same information about
the system, so even if final state $e$ does not know precisely which
environmental path was taken, it has still retained all information
that was gathered about the system (see Thm \ref{thm:O=00005B=00005D make-up}). 

Turning this around, let's assume that $\Theta_{e}\neq(\mathcal{S}\cdot\Theta_{e})\times(\mathcal{E}\cdot\Theta_{e})$.
There are then $\bar{e}[0,\lambda],\bar{e}^{\prime}[0,\lambda^{\prime}]\in\mathcal{E}\cdot\Theta_{e}$
that are paired with different sets of system paths. In this case,
the final state has forgotten information about the system that it
gathered as it passed along one of these paths. This leads us to define
recorders that never loose information about the system as:
\begin{defn}
A recorder is \emph{all-reet} if for every $e\in\mathcal{F}$, $\Theta_{e}=(\mathcal{S}\cdot\Theta_{e})\times(\mathcal{E}\cdot\Theta_{e})$.
\end{defn}

An all-reet recorder is all retaining; once an all-reet recorder records
a bit of information about the system, it never forgets it. The next
theorem enumerates some useful properties of all-reet recorders:
\begin{thm}
\label{thm:All-Reet props}If $Z$ is an all-reet recorder then:

1) For every $e\in\mathcal{F}$, every $\bar{s}_{1}[0,\lambda_{1}],\bar{s}_{2}[0,\lambda_{2}]\in\mathcal{O}_{e}$,
$\lambda_{1}=\lambda_{2}$

2) For every $p\in Int(Z)$, every $\bar{s}_{1}[0,\lambda_{1}],\bar{s}_{2}[0,\lambda_{2}]\in\Sigma_{p}$,
$\lambda_{1}=\lambda_{2}$

3) Assume $\bar{s}[0,\lambda_{e}]\in\mathcal{O}_{e}$. For every $e\in\mathcal{F}$,
every $\lambda\in[0,\lambda_{e}]$, $\mathcal{O}_{e}=\mathcal{O}_{e}[0,\lambda]\circ\mathcal{O}_{e}[\lambda,\lambda_{e}]$ 
\end{thm}

\begin{proof}
1) Immediate from $\Theta_{e}=\mathcal{O}_{e}\times(\mathcal{E}\cdot\Theta_{e})$

2) Follows from (1), and the fact that $\mathbb{D}$ is a homogeneous
dynamic space

3) $\Theta_{e}=\mathcal{O}_{e}\times(\mathcal{E}\cdot\Theta_{e})$,
so if $\bar{s}_{1}[0,\lambda],\bar{s}_{2}[0,\lambda]\in\mathcal{O}_{e}$
there's a $\bar{e}[0,\lambda]\in\mathcal{E}\cdot\Theta_{e}$ s.t.
$(\bar{s}_{1}[0,\lambda],\bar{e}[0,\lambda]),(\bar{s}_{2}[0,\lambda],\bar{e}[0,\lambda])\in\Theta_{e}$.
Since $\mathbb{D}$ is a dynamic space, if $\bar{s}_{1}[0,\lambda](\lambda^{\prime})=\bar{s}_{2}[0,\lambda](\lambda^{\prime})$,
$(\bar{s}_{1}[0,\lambda^{\prime}]\circ\bar{s}_{2}[\lambda^{\prime},\lambda],\bar{e}[0,\lambda])\in\Theta_{e}$.
\end{proof}
Using part 1 of the above theorem, for any outcome, $\mathcal{O}_{e}$,
of an all-reet recorder, we can define $\Lambda(\mathcal{O}_{e})$
so that if $\bar{s}[0,\lambda]\in\mathcal{O}_{e}$ then $\Lambda(\mathcal{O}_{e})\equiv\lambda$.
Part 3 could then be written: For every $e\in\mathcal{F}$, every
$\lambda\in[0,\Lambda(\mathcal{O}_{e})]$, $\mathcal{O}_{e}=\mathcal{O}_{e}[0,\lambda]\circ\mathcal{O}_{e}[\lambda,\Lambda(\mathcal{O}_{e})]$.

\subsubsection{Boolean Recorders}

The most obvious example of the blurring of intrinsic and extrinsic
uncertainty occurs when recorders decide coverings that are not partitions.
To complete the disentanglement of intrinsic and extrinsic uncertainty,
we therefore consider the property that ensures recorders decide partitions.
\begin{defn}
A recorder is \emph{Boolean} if for any $e,e^{\prime}\in\mathcal{F}$,
either for all $\bar{s}[0,\lambda]\in\mathcal{O}_{e}$, $\bar{s}^{\prime}[0,\lambda^{\prime}]\in\mathcal{O}_{e^{\prime}}$,
$\lambda_{1}\equiv Min(\lambda,\lambda^{\prime})$, $\bar{s}[0,\lambda_{1}]\neq\bar{s}^{\prime}[0,\lambda_{1}]$,
or for all $\bar{s}[0,\lambda]\in\mathcal{O}_{e}\bigtriangleup\mathcal{O}_{e^{\prime}}$
there's a $\lambda^{\prime}<\lambda$ s.t. $\bar{s}[0,\lambda^{\prime}]\in\mathcal{O}_{e}\bigcap\mathcal{O}_{e^{\prime}}$.
\end{defn}

(``$\mathcal{O}_{e}\bigtriangleup\mathcal{O}_{e^{\prime}}$'' is
the symmetric difference: $\mathcal{O}_{e}\bigtriangleup\mathcal{O}_{e^{\prime}}=(\mathcal{O}_{e}\setminus\mathcal{O}_{e^{\prime}})\bigcup(\mathcal{O}_{e^{\prime}}\setminus\mathcal{O}_{e})$.)

The significance of Boolean recorders is captured in the following
theorem:
\begin{thm}
\label{thm:If boolean decidable then partition} 1) If recorder $Z$
is Boolean, and it decides covering $K$, then $K$ is a partition.

2) A recorder decides a partition for any system dynamic set iff it
is Boolean.
\end{thm}

\begin{proof}
1) Take any $e,e^{\prime}\in\mathcal{F}$. If for all $\bar{s}[0,\lambda_{1}]\in\mathcal{O}_{e}$,
$\bar{s}^{\prime}[0,\lambda_{2}]\in\mathcal{O}_{e^{\prime}},$ $\lambda_{3}\equiv Min(\lambda_{1},\lambda_{2})$,
$\bar{s}[0,\lambda_{3}]\neq\bar{s}^{\prime}[0,\lambda_{3}]$, then
for all $\lambda\in\varLambda_{\mathbb{D}}$, $S_{\rightarrow\mathcal{O}_{e}^{\lambda}\rightarrow}\bigcap S_{\rightarrow\mathcal{O}_{e^{\prime}}^{\lambda}\rightarrow}=\emptyset$.
If for all $\bar{s}[0,\lambda_{1}]\in\mathcal{O}_{e}\bigtriangleup\mathcal{O}_{e^{\prime}}$
there's a $\lambda_{0}<\lambda_{1}$ s.t. $\bar{s}[0,\lambda_{0}]\in\mathcal{O}_{e}\bigcap\mathcal{O}_{e^{\prime}}$
then $S_{\rightarrow\mathcal{O}_{e}^{\lambda}\rightarrow}=S_{\rightarrow\mathcal{O}_{e^{\prime}}^{\lambda}\rightarrow}$
for all $\lambda\in\varLambda_{\mathbb{D}}$.

2) A recorder is not Boolean if and only if for some $e,e^{\prime}\in\mathcal{F}$,
and some $\bar{s}[0,\lambda_{1}]\in\mathcal{O}_{e}$, $\bar{s}^{\prime}[0,\lambda_{2}]\in\mathcal{O}_{e^{\prime}}$
($\lambda_{1}<\lambda_{2}$), $\bar{s}[0,\lambda_{1}]=\bar{s}^{\prime}[0,\lambda_{1}]$
and $\bar{s}[0,\lambda_{1}]\notin\mathcal{O}_{e^{\prime}}$. Under
these conditions $S_{\rightarrow\mathcal{O}_{e}\rightarrow}\bigcap S_{\rightarrow\mathcal{O}_{e^{\prime}}\rightarrow}=\emptyset$
requires $S_{\rightarrow\bar{s}^{\prime}[0,\lambda_{2}]\rightarrow}=\emptyset$,
while $S_{\rightarrow\mathcal{O}_{e}\rightarrow}=S_{\rightarrow\mathcal{O}_{e^{\prime}}\rightarrow}$
requires that for every $\bar{s}_{1}\in S_{\rightarrow\bar{s}[0,\lambda_{1}]\rightarrow}$,
there's a $\lambda_{3}>\lambda_{1}$ s.t $\bar{s}_{1}[0,\lambda_{3}]\in\mathcal{O}_{e^{\prime}}$.
This clearly cannot hold for all dynamic sets. For example, there
will always be dynamic sets s.t. $S_{\rightarrow\bar{s}^{\prime}[0,\lambda_{2}]\rightarrow}\neq\emptyset$,
and for all $\lambda>\lambda_{1}$, there's a $\lambda^{\prime}\in(\lambda_{1},\lambda]$
s.t. for some $s\in S_{\rightarrow\bar{s}[0,\lambda_{1}]\rightarrow}(\lambda^{\prime})$,
$s\notin\mathcal{O}_{e^{\prime}}(\lambda^{\prime})$.
\end{proof}
As the next theorem shows, the conditions under which a recorder is
Boolean are simplified when the recorder is all-reet:
\begin{thm}
\label{thm:When Boolean}If a recorder is all-reet, it is Boolean
if and only if the following two conditions hold:

1) For every $e,e^{\prime}\in\mathcal{F}$ either $\mathcal{O}_{e}=\mathcal{O}_{e^{\prime}}$
or $\mathcal{O}_{e}\bigcap\mathcal{O}_{e^{\prime}}=\emptyset$

2) For every $\bar{s}_{1}[0,\lambda_{1}],\,\bar{s}_{2}[0,\lambda_{2}]\in\mathcal{O}_{\mathcal{F}}$
s.t. $\lambda_{1}<\lambda_{2}$, $\bar{s}_{1}[0,\lambda_{1}]\neq\bar{s}_{2}[0,\lambda_{1}]$.
\end{thm}

The first condition says that the set of outcomes partition $\mathcal{O}_{\mathcal{F}}$,
and the second says that when an experiment terminates depends entirely
on what occurs in the system. One would likely expect any experiment
to satisfy both conditions.
\begin{proof}
First, we'll show that if the above two conditions hold, then the
recorder is Boolean.

If $\mathcal{O}_{e}=\mathcal{O}_{e^{\prime}}$ then $\mathcal{O}_{e}\bigtriangleup\mathcal{O}_{e^{\prime}}=\emptyset$,
so ``for all $\bar{s}[0,\lambda]\in\mathcal{O}_{e}\bigtriangleup\mathcal{O}_{e^{\prime}}$
there's a $\lambda^{\prime}<\lambda$ s.t. $\bar{s}[0,\lambda^{\prime}]\in\mathcal{O}_{e}\bigcap\mathcal{O}_{e^{\prime}}$''
holds trivially.

If $\mathcal{O}_{e}\bigcap\mathcal{O}_{e^{\prime}}=\emptyset$ and
for every $\bar{s}_{1}[0,\lambda_{1}],\,\bar{s}_{2}[0,\lambda_{2}]\in\mathcal{O}_{\mathcal{F}}$
s.t. $\lambda_{1}<\lambda_{2}$, $\bar{s}_{1}[0,\lambda_{1}]\neq\bar{s}_{2}[0,\lambda_{1}]$,
then for every $\bar{s}[0,\lambda]\in\mathcal{O}_{e}$, $\bar{s}^{\prime}[0,\lambda^{\prime}]\in\mathcal{O}_{e^{\prime}}$,
with $\lambda_{1}\equiv Min(\lambda,\lambda^{\prime})$, $\bar{s}[0,\lambda_{1}]\neq\bar{s}^{\prime}[0,\lambda_{1}]$.

To prove the two conditions, we need only apply Thm \ref{thm:All-Reet props}.1
to the Boolean condition.

If $\Lambda(\mathcal{O}_{e})=\Lambda(\mathcal{O}_{e^{\prime}})$,
then the Boolean condition reduces to: either $\mathcal{O}_{e}=\mathcal{O}_{e^{\prime}}$
or $\mathcal{O}_{e}\bigcap\mathcal{O}_{e^{\prime}}=\emptyset$.

If $\Lambda(\mathcal{O}_{e})<\Lambda(\mathcal{O}_{e^{\prime}})$ then
$\mathcal{O}_{e}\bigcap\mathcal{O}_{e^{\prime}}=\emptyset$ and, by
the Boolean condition, for all $\bar{s}[0,\lambda]\in\mathcal{O}_{e}$,
$\bar{s}^{\prime}[0,\lambda^{\prime}]\in\mathcal{O}_{e^{\prime}}$,
$\bar{s}[0,\lambda]\neq\bar{s}^{\prime}[0,\lambda]$.
\end{proof}
The second condition in the prior theorem has a useful application,
it helps to clarify the weak unbiased condition. As it stands, the
weak unbiased condition is convoluted and difficult to apply. The
above theorem allows us to extract the following simple property that
is central to its meaning:
\begin{thm}
\label{thm:The Unbiased Thm}If recorder $Z$ is all-reet and Boolean
then for all $p_{1},p_{2}\in Int(Z)$, if $\Sigma_{p_{1}}\bigcap\Sigma_{p_{2}}\neq\emptyset$
then $\Sigma_{p_{1}\rightarrow}=\Sigma_{p_{2}\rightarrow}$
\end{thm}

\begin{proof}
Since $\Sigma_{p_{1}}\bigcap\Sigma_{p_{2}}\neq\emptyset$, $p_{1}$
and $p_{2}$ are undifferentiated, so the theorem follows from the
weak unbiased condition, together with the second condition in Thm
\ref{thm:When Boolean}.
\end{proof}

\subsubsection{Ideal Recorders}
\begin{defn}
A recorder is \emph{ideal} if it is Boolean and all-reet.
\end{defn}

Ideal recorders lack the ambiguities described at the beginning of
this section, and so provide our preferred formalization of experiment.
It should be stressed that all the axioms for ideal recorders have
been motivated solely by the requirements of the experimental method,
and not by the dictates of any particular physical theory. This means
that a universe that does not allow the axioms for recorders to be
realized is one in which the scientific method cannot be applied,
and a universe that does not allow the axioms for \emph{ideal} recorders
to be realized is one in which experimental error cannot be handled
systematically. However, this does not mean that ideal recorders necessarily
have to be constructible within an experimentally verifiable physical
theory. In particular, a theory may describe systems being experimented
on, without being able to describe the closed system that includes
both the system being experimented on, and the experimental equipment.
``Dynamic probability spaces'', to be described in some detail in
the next part, provide a template for such theories.

We now turn to the question of what experiments are capable of telling
us. This entails describing the sorts of outcomes that experiments
can produce. For ideal recorders, the central theorem is: If $Z$
is an ideal recorder, then for any $e\in\mathcal{F}$, $\lambda\in[0,\Lambda(\mathcal{O}_{e})]$,
$\mathcal{O}_{e}=|\mathcal{O}_{e}[0,\lambda]|\circ\mathcal{O}_{e}[\lambda,\Lambda(\mathcal{O}_{e})]$.

Before proving this assertion, we note its plausibility. Just as system
paths can be partitioned based on experimental outcomes, they can
be partitioned based on the what the experiment has learned as of
any time, $\lambda$. For all-reet recorders, these partitions are
composed of the $|\mathcal{O}_{e}[0,\lambda]|$'s (see Defn \ref{def: |O=00005B=00005D|}).
The assertion therefore states that an ideal recorder's outcomes are
the concatenation of what has been learned as of $\lambda$ with what
is learned after $\lambda$.

The remainder of this section is dedicated to establishing this result,
starting with the following theorem regarding all-reet recorders:
\begin{thm}
\label{thm:O=00005B=00005D make-up}If $Z$ is an all-reet recorder
then for every $e\in\mathcal{F}$, every $\lambda\in[0,\Lambda(\mathcal{O}_{e})]$,
every $e_{\lambda}\in\mathcal{E}\cdot\Theta_{e}(\lambda)$, $\mathcal{O}_{e}[0,\lambda]=\bigcup_{s\in\mathcal{O}_{e}(\lambda)}\Sigma_{(s,e_{\lambda})}$
\end{thm}

\begin{proof}
First, to show that $\bigcup_{s\in\mathcal{O}_{e}(\lambda)}\Sigma_{(s,e_{\lambda})}\subset\mathcal{O}_{e}[0,\lambda]$.
Take any $s_{\lambda}\in\mathcal{O}_{e}(\lambda)$. Since $Z$ is
all-reet, $(s_{\lambda},e_{\lambda})\in\Theta_{e}(\lambda)$. Take
any $\bar{s}[0,\lambda]\in\Sigma_{(s_{\lambda},e_{\lambda})}$, and
any $\bar{e}[0,\lambda]\in\mathcal{E}\cdot\omega_{(s_{\lambda},e_{\lambda})}$.
Once again, because $Z$ is all-reet, $(\bar{s}[0,\lambda],\bar{e}[0,\lambda])\in\omega_{(s_{\lambda},e_{\lambda})}$.
Because $(s_{\lambda},e_{\lambda})\in\Theta_{e}(\lambda)$, there
must be a $\bar{p}[0,\lambda^{\prime}]\in\omega_{e}$ s.t. $\bar{p}[0,\lambda^{\prime}](\lambda)=(s_{\lambda},e_{\lambda})$.
Because $\mathbb{D}$ is a dynamic space, $(\bar{s}[0,\lambda],\bar{e}[0,\lambda])\circ\bar{p}[\lambda,\lambda^{\prime}]\in\omega_{e}$.
It is easy to see that for all $0\leq\lambda^{\prime\prime}<\lambda$,
$\bar{e}[0,\lambda](\lambda^{\prime\prime})\notin\mathcal{F}$, so
$(\bar{s}[0,\lambda],\bar{e}[0,\lambda])\in\Theta_{e}[0,\lambda]$,
and $\bar{s}[0,\lambda]\in\mathcal{O}_{e}[0,\lambda]$.

Next, to show that $\mathcal{O}_{e}[0,\lambda]\subset\bigcup_{s\in\mathcal{O}_{e}(\lambda)}\Sigma_{(s,e_{\lambda})}$.
If $\bar{s}[0,\lambda]\in\mathcal{O}_{e}[0,\lambda]$ then, because
$Z$ is all-reet, for any $\bar{e}[0,\lambda]\in\mathcal{E}\cdot\Theta_{e}[0,\lambda]$,
$(\bar{s}[0,\lambda],\bar{e}[0,\lambda])\in\Theta_{e}[0,\lambda]$.
Take any $\bar{e}^{\prime}[0,\lambda]\in\mathcal{E}\cdot\Theta_{e}[0,\lambda]$
s.t. $\bar{e}^{\prime}[0,\lambda](\lambda)=e_{\lambda}$; $(\bar{s}[0,\lambda],\bar{e}^{\prime}[0,\lambda])\in\Theta_{e}[0,\lambda]$.
With $s\equiv\bar{s}[0,\lambda](\lambda)$ , it follows that $\bar{s}[0,\lambda]\in\Sigma_{(s,e_{\lambda})}$.
\end{proof}
The next theorem lists some some of the consequences of Thm \ref{thm:O=00005B=00005D make-up}.
\begin{defn}
For any recorder, any $e\in\mathcal{F}$, $[e]\equiv\{e^{\prime}\in\mathcal{F}:\mathcal{O}_{e}=\mathcal{O}_{e^{\prime}}\}$
\end{defn}

\begin{cor}
\label{thm:Corollaries to O=00005B=00005D make-up}If $Z$ is an all-reet
recorder, $p,p_{1},p_{2}\in Int(Z)$, $e\in\mathcal{F}$, and $\lambda\in[0,\Lambda(\mathcal{O}_{e})]$

1) If $(s_{1},e_{1}),(s_{2},e_{2})\in\Theta_{e}(\lambda)$ then $(s_{1},e_{2})\in\Theta_{e}(\lambda)$ 

2) If $(s,e_{1}),(s,e_{2})\in\Theta_{e}(\lambda)$ then $\Sigma_{(s,e_{1})}=\Sigma_{(s,e_{2})}$

3) If $(s,e_{1}),(s,e_{2})\in\Theta_{[e]}(\lambda)$ then $\Sigma_{(s,e_{1})}=\Sigma_{(s,e_{2})}$
\end{cor}

\begin{proof}
1 and 2 are immediate from Thm \ref{thm:O=00005B=00005D make-up}.
3 follows from Thm \ref{thm:O=00005B=00005D make-up} and the definition
of $[e]$. 
\end{proof}
We now turn to ideal recorders.
\begin{thm}
\label{thm:sigma in an iea}For ideal recorder $Z$, for any $p_{1},p_{2}\in Int(Z)$ 

1) If $\mathcal{S}\cdot p_{1}=\mathcal{S}\cdot p_{2}$, and for some
$e\in\mathcal{F}$, $\lambda\in[0,\Lambda(\mathcal{O}_{e})]$, $p_{1},p_{2}\in\Theta_{[e]}(\lambda)$
then $\Sigma_{p_{1}}=\Sigma_{p_{2}}$, in all other cases $\Sigma_{p_{1}}\bigcap\Sigma_{p_{2}}=\emptyset$.

2) If $\Sigma_{p_{1}}=\Sigma_{p_{2}}$ then for all $e\in\mathcal{F}$,
$\lambda\in[0,\Lambda(\mathcal{O}_{e})]$, $p_{1}\in\Theta_{[e]}(\lambda)$
iff $p_{2}\in\Theta_{[e]}(\lambda)$ 
\end{thm}

\begin{proof}
1) If the conditions hold then $\Sigma_{p_{1}}=\Sigma_{p_{2}}$ by
Thm \ref{thm:Corollaries to O=00005B=00005D make-up}.3.

If $p_{1}\cdot\mathcal{S}\neq p_{2}\cdot\mathcal{S}$ then clearly
$\Sigma_{p_{1}}\bigcap\Sigma_{p_{2}}=\emptyset$.

If there's doesn't exist an $e\in\mathcal{F}$ s.t. $p_{1},p_{2}\in Ran(\Theta_{[e]})$
then, since $Z$ is Boolean, $\Sigma_{p_{1}}\circ\Sigma_{p_{1}\rightarrow}\bigcap\Sigma_{p_{2}}\circ\Sigma_{p_{2}\rightarrow}=\emptyset$.
By Thm \ref{thm:The Unbiased Thm}, $\Sigma_{p_{1}}\bigcap\Sigma_{p_{2}}=\emptyset$ 

If there exits an $e\in\mathcal{F}$ s.t. $p_{1},p_{2}\in Ran(\Theta_{[e]})$,
but no $\lambda\in[0,\Lambda(\mathcal{O}_{e})]$ s.t. $p_{1},p_{2}\in\Theta_{[e]}(\lambda)$
then $\Sigma_{p_{1}}\bigcap\Sigma_{p_{2}}=\emptyset$ by Thm \ref{thm:All-Reet props}.2.

2) By Thm \ref{thm:The Unbiased Thm} $\Sigma_{p_{1}}\circ\Sigma_{p_{1}\rightarrow}=\Sigma_{p_{2}}\circ\Sigma_{p_{2}\rightarrow}$;
since $Z$ is Boolean $p_{1}\in Ran(\Theta_{[e]})$ iff $p_{2}\in Ran(\Theta_{[e]})$.
From Thm \ref{thm:All-Reet props}.2 it then follows that $p_{1}\in\Theta_{[e]}(\lambda)$
iff $p_{2}\in\Theta_{[e]}(\lambda)$.
\end{proof}
\begin{thm}
\label{thm:|O=00005B=00005D| make-up}If $Z$ is an ideal recorder
then for every $e\in\mathcal{F}$, every $\lambda\in[0,\Lambda(\mathcal{O}_{e})]$,
every $e_{\lambda}\in\mathcal{E}\cdot\Theta_{[e]}(\lambda)$, $|\mathcal{O}_{e}[0,\lambda]|=\bigcup_{s\in|\mathcal{O}_{e}[0,\lambda]|(\lambda)}\Sigma_{(s,e_{\lambda})}$
\end{thm}

\begin{proof}
A: If $Z$ is an ideal recorder then for every $e\in\mathcal{F}$,
every $\lambda\in[0,\Lambda(\mathcal{O}_{e})]$, every $e_{\lambda}\in\mathcal{E}\cdot\Theta_{[e]}(\lambda)$,
$\mathcal{O}_{e}[0,\lambda]=\bigcup_{s\in\mathcal{O}_{e}(\lambda)}\Sigma_{(s,e_{\lambda})}$

- Immediate from Thm \ref{thm:O=00005B=00005D make-up} and Thm \ref{thm:Corollaries to O=00005B=00005D make-up}.3
-

It is sufficient to show that for every $n\in\mathbb{N},$ $|\mathcal{O}_{e}[0,\lambda]|_{n}=\bigcup_{s\in|\mathcal{O}_{e}[0,\lambda]|_{n}(\lambda)}\Sigma_{(s,e_{\lambda})}$.
By (A) this holds for $n=0$. Assume it holds for $n=i$ and consider
$n=i+1$.

For any $\mathcal{O}_{e^{\prime}}$ s.t. $\mathcal{O}_{e^{\prime}}[0,\lambda]\bigcap|\mathcal{O}_{e}[0,\lambda]|_{i}\neq\emptyset$
for every $e_{2}\in\mathcal{E}\cdot\Theta_{[e^{\prime}]}(\lambda)$,
$\mathcal{O}_{e^{\prime}}[0,\lambda]=\bigcup_{s\in\mathcal{O}_{e^{\prime}}(\lambda)}\Sigma_{(s,e_{2})}$.
By assumption $|\mathcal{O}_{e}[0,\lambda]|_{i}=\bigcup_{s\in|\mathcal{O}_{e}[0,\lambda]|_{i}(\lambda)}\Sigma_{(s,e_{\lambda})}$,
so by Thm \ref{thm:sigma in an iea}.1 there's an $s\in|\mathcal{O}_{e}[0,\lambda]|_{i}(\lambda)\bigcap\mathcal{O}_{e^{\prime}}(\lambda)$
s.t. $\Sigma_{(s,e_{2})}=\Sigma_{(s,e_{\lambda})}$; by Thm \ref{thm:sigma in an iea}.2
$(s,e_{\lambda})\in\Theta_{[e^{\prime}]}(\lambda)$, so by (A) $\mathcal{O}_{e^{\prime}}[0,\lambda]=\bigcup_{s\in\mathcal{O}_{e^{\prime}}(\lambda)}\Sigma_{(s,e_{\lambda})}$
. Since $|\mathcal{O}_{e}[0,\lambda]|_{i+1}$is the union over all
such $\mathcal{O}_{e^{\prime}}[0,\lambda]$ it follows immediately
that $|\mathcal{O}_{e}[0,\lambda]|_{i+1}=\bigcup_{s\in|\mathcal{O}_{e}[0,\lambda]|_{i+1}(\lambda)}\Sigma_{(s,e_{\lambda})}$.
\end{proof}
Which brings us to the desired result:
\begin{thm}
\label{thm:ip establishment thm}If $Z$ is an ideal recorder then
for any $e\in\mathcal{F}$, $\lambda\in[0,\Lambda(\mathcal{O}_{e})]$,
$\mathcal{O}_{e}=|\mathcal{O}_{e}[0,\lambda]|\circ\mathcal{O}_{e}[\lambda,\Lambda(\mathcal{O}_{e})]$
\end{thm}

\begin{proof}
With $e_{\lambda}\in\mathcal{E}\cdot\Theta_{e}(\lambda)$:

$\mathcal{O}_{e}=\mathcal{O}_{e}[0,\lambda]\circ\mathcal{O}_{e}[\lambda,\Lambda(\mathcal{O}_{e})]$
(Thm \ref{thm:All-Reet props}.3)

$=(\bigcup_{s\in\mathcal{O}_{e}(\lambda)}\Sigma_{(s,e_{\lambda})})\circ\mathcal{O}_{e}[\lambda,\Lambda(\mathcal{O}_{e})]$
(Thm \ref{thm:O=00005B=00005D make-up})

$=(\bigcup_{s\in|\mathcal{O}_{e}[0,\lambda]|(\lambda)}\Sigma_{(s,e_{\lambda})})\circ\mathcal{O}_{e}[\lambda,\Lambda(\mathcal{O}_{e})]$
(any path segment ending with an $s\in|\mathcal{O}_{e}[0,\lambda]|(\lambda)\setminus\mathcal{O}_{e}(\lambda)$
will not be concatenated with $\mathcal{O}_{e}[\lambda,\Lambda(\mathcal{O}_{e})]$)

$=|\mathcal{O}_{e}[0,\lambda]|\circ\mathcal{O}_{e}[\lambda,\Lambda(\mathcal{O}_{e})]$
(Thm \ref{thm:|O=00005B=00005D| make-up})
\end{proof}

\subsection{Ideal Partitions}

\subsubsection{Ideal Recorders \& Ideal Partitions}

Thm \ref{thm:ip establishment thm} allows us to identify the coverings
that ideal recorders are capable of deciding:
\begin{defn}
A partition, $\gamma$, of dynamic set, $S$, is \emph{ideal} if 

1) For all $\alpha\in\gamma$, all $\lambda\in\Lambda_{S}$, $\alpha=|\alpha[-\infty,\lambda]|\circ\alpha[\lambda,\infty]$

2) There exists a $\lambda_{0}\in\Lambda_{S}$ s.t. for all $\alpha\in\gamma$,
all $\lambda\leq\lambda_{0}$, $\alpha=S[-\infty,\lambda]\circ\alpha[\lambda,\infty]$.

3) For all $\alpha\in\gamma$, there exists a $\lambda_{\alpha}\in\Lambda_{S}$
s.t. $\alpha=S_{\alpha[-\infty,\lambda_{\alpha}]\rightarrow}$.
\end{defn}

The first condition reflects Thm \ref{thm:ip establishment thm},
while the next two reflect Thm \ref{thm:trivial}. The following three
theorems establish the relationship between ideal recorders and ideal
partitions.
\begin{thm}
\label{thm:if decided then ip}If a covering, $K$, of dynamic set
$S$ is properly decided by an ideal recorder, $Z$, then $K$ is
an ideal partition
\end{thm}

\begin{proof}
Because $K$ is properly decided by $Z$, there must be a $\lambda_{0}\in\Lambda_{S}$
s.t. $K\equiv\{S_{\rightarrow\mathcal{O}_{e}^{\lambda_{0}}\rightarrow}:e\in\mathcal{F}\:and\:S_{\rightarrow\mathcal{O}_{e}^{\lambda_{0}}\rightarrow}\neq\emptyset\}$,
and $S$ is unbiased with respect to $(\lambda_{0},K)$. For each
$\alpha=S_{\rightarrow\mathcal{O}_{e}^{\lambda_{0}}\rightarrow}\in K$
define $\lambda_{\alpha}\equiv\lambda_{0}+\Lambda(\mathcal{O}_{e})$.

A: $K$ is a partition

- Immediate from Thm \ref{thm:If boolean decidable then partition}
-

B: For all $\alpha\in K$, $\lambda\leq\lambda_{0}$, $\alpha=S[-\infty,\lambda]\circ\alpha[\lambda,\infty]$

- Follows from $\alpha=S_{\rightarrow\mathcal{O}_{e}^{\lambda_{0}}\rightarrow}$
and the fact that for all $\lambda\leq\lambda_{0}$, $S=S[-\infty,\lambda]\circ S[\lambda,\infty]$

C: For every $\alpha\in K$, $\alpha=S_{\alpha[-\infty,\lambda_{\alpha}]\rightarrow}$

- Follows from $\alpha=S_{\rightarrow\mathcal{O}_{e}^{\lambda_{0}}\rightarrow}$
and $\lambda_{\alpha}=\lambda_{0}+\Lambda(\mathcal{O}_{e})$ -

It remains to show that for all $\alpha\in K$, $\lambda\in\Lambda_{S}$,
$\alpha=|\alpha[-\infty,\lambda]|\circ\alpha[\lambda,\infty]$. As
$\alpha\subset|\alpha[-\infty,\lambda]|\circ\alpha[\lambda,\infty]$,
it only needs to be shown that $|\alpha[-\infty,\lambda]|\circ\alpha[\lambda,\infty]\subset\alpha$.
To do so, it will be helpful to first establish the following:

D: For all $\lambda>\lambda_{0}$, $\alpha\in K$, $|\alpha[-\infty,\lambda]|=S[-\infty,\lambda_{0}]\circ|\alpha[\lambda_{0},\lambda]|$

- It is sufficient to show that for all $n\in\mathbb{N}$, $|\alpha[-\infty,\lambda]|_{n}=S[-\infty,\lambda_{0}]\circ|\alpha[\lambda_{0},\lambda]|_{n}$.
By (B), this holds for $n=0$. Assume it holds for $n=i$. By (B)
and assumption on $i$, for all $\beta\in K$, $\beta[-\infty,\lambda]\bigcap|\alpha[-\infty,\lambda]|_{i}\neq\emptyset$
iff $\beta[\lambda_{0},\lambda]\bigcap|\alpha[\lambda_{0},\lambda]|_{i}\neq\emptyset$.
Since $|\alpha[x,\lambda]|_{i+1}$ is equal to the union over the
$\beta$'s that intersect $|\alpha[x,\lambda]|_{i}$, and because
$|\alpha[-\infty,\lambda]|_{i}$ \& $|\alpha[\lambda_{0},\lambda]|_{i}$
are intersected by the same set of $\beta\in K$, $|\alpha[-\infty,\lambda]|_{i+1}=S[-\infty,\lambda_{0}]\circ|\alpha[\lambda_{0},\lambda]|_{i+1}$
-

Now to complete the proof by showing that $|\alpha[-\infty,\lambda]|\circ\alpha[\lambda,\infty]\subset\alpha$

For $\lambda\leq\lambda_{0}$: Immediate from (B).

For $\lambda\in(\lambda_{0},\lambda_{\alpha})$: Choose any $\bar{p}_{1}[-\infty,\lambda]\in|\alpha[-\infty,\lambda]|$
and $\bar{p}_{2}[\lambda,\infty]\in\alpha[\lambda,\infty]$. Since
$S$ is unbiased with respect to $(\lambda_{0},K)$, $\bar{p}_{1}[-\infty,\lambda]\circ\bar{p}_{2}[\lambda,\infty]\in S$;
it remains to show that it is an element of $\alpha$. It follows
from (D) that $\bar{p}_{1}[\lambda_{0},\lambda]\in|\alpha[\lambda_{0},\lambda]|$,
and so from Thm \ref{thm:ip establishment thm} that $\bar{p}_{1}[\lambda_{0},\lambda]\circ\bar{p}_{2}[\lambda,\lambda_{\alpha}]\in\alpha[\lambda_{0},\lambda_{\alpha}]$.
It then follows from (B) that $\bar{p}_{1}[-\infty,\lambda]\circ\bar{p}_{2}[\lambda,\lambda_{\alpha}]\in\alpha[-\infty,\lambda_{\alpha}]$.
Since $\alpha=S_{\alpha[-\infty,\lambda_{\alpha}]\rightarrow}$, and
$\bar{p}_{1}[-\infty,\lambda]\circ\bar{p}_{2}[\lambda,\infty]\in S$,
$\bar{p}_{1}[-\infty,\lambda]\circ\bar{p}_{2}[\lambda,\infty]\in\alpha$.

For $\lambda\geq\lambda_{\alpha}$: Because $S$ is unbiased with
respect to $(\lambda_{0},K)$, $\alpha[-\infty,\lambda]\circ\alpha[\lambda,\infty]\subset S$,
and so by (C) $\alpha=\alpha[-\infty,\lambda]\circ\alpha[\lambda,\infty]$.
By (A) \& (C), $|\alpha[-\infty,\lambda]|=\alpha[-\infty,\lambda]$,
giving us $\alpha=|\alpha[-\infty,\lambda]|\circ\alpha[\lambda,\infty]$. 
\end{proof}
The inverse can also be established, that all ideal partitions are
decided by ideal recorders. Note, however, that the ideal recorder
may not exist within a particular dynamic space. What is shown is
that for any ideal partition, an ideal recorder can be constructed
that decides it.
\begin{thm}
\label{thm:ip's are ideally decidable}Every ideal partition is properly
decided by an ideal recorder
\end{thm}

\begin{proof}
Assume $\gamma$ is an ideal partition of dynamic set $S$. We will
describe an ideal recorder, $Z$, that decides $\gamma$.

1) Start by constructing dynamic space $D_{0}=S\times E_{0}$ as follows:
For every $\lambda\in\Lambda_{S}$ create a set $E(\lambda),$ and
a bijection $b_{\lambda}:|\gamma[-\infty,\lambda]|\rightarrow E(\lambda)$
s.t. for all $\lambda_{1}\neq\lambda_{2}$, $E(\lambda_{1})\bigcap E(\lambda_{2})=\emptyset$;
$(\bar{s},\bar{e})\in D_{0}$ iff $\bar{s}\in S$ and with $\bar{s}\in\alpha\in\gamma$,
for all $\lambda\in\Lambda$, $\bar{e}(\lambda)=b_{\lambda}(|\alpha[-\infty,\lambda]|)$

2) From $D_{0}$ construct $Z$'s dynamic space, $\mathbb{D}$: 

If $\Lambda_{S}$ is unbounded from below, for every $\lambda\in\Lambda_{S}$
define $D_{\lambda}\equiv\{\bar{p}:for\:some\:\bar{p}^{\prime}\in D_{0},\:for\:all\:\lambda^{\prime}\in\Lambda_{S},\:\bar{p}(\lambda^{\prime})=\bar{p}^{\prime}(\lambda^{\prime}+\lambda)\}$.
That is, $D_{\lambda}$ is $D_{0}$ shifted by $\lambda$. $\mathbb{D}\equiv\bigcup_{\lambda\in\Lambda}D_{\lambda}$.

If $\Lambda_{S}$ is bounded from below, take any dynamic set, $S^{*}$,
s.t. $\Lambda_{S_{0}^{*}}$ is unbounded from below, $S_{0}^{*}[0,\infty]=S_{0}$,
and for all $\lambda\leq0$, $S^{*}=S^{*}[-\infty,\lambda]\circ S^{*}[\lambda,\infty]$.
Construct $E_{0}^{*}$ and $D_{0}^{*}=S^{*}\times E_{0}^{*}$ as in
the $1^{st}$ step (each element of $\gamma$ being extended to $S^{*}$).
Construct $\mathbb{D}^{*}$ from $D_{0}^{*}$ as above. Take take
$\mathbb{D}=\mathbb{D}^{*}[0,\infty]$.

3) $I$: Select any $\lambda_{0}\in\Lambda_{\mathbb{D}}$ s.t. for
all $\lambda\leq\lambda_{0}$, all $\alpha\in\gamma$, $\alpha=S[-\infty,\lambda]\circ\alpha[\lambda,\infty]$.
$I\equiv D_{0}(\lambda_{0})$

4) $F$: For every $\alpha\in\gamma$ select a $\lambda_{\alpha}\in\Lambda_{\mathbb{D}}$
s.t. $\alpha=S_{\alpha[-\infty,\lambda_{\alpha}]\rightarrow}$. $F\equiv\bigcup_{\alpha\in\gamma}\alpha(\lambda_{\alpha})\times\{b_{\lambda_{a}}(|\alpha[-\infty,\lambda_{\alpha}]|)\}$.

Now to show that $Z$ is an ideal recorder that properly decides $\gamma$.

A: $D_{0}$ is a dynamic space

- Take any $(\bar{s}_{1},\bar{e}_{1}),(\bar{s}_{2},\bar{e}_{2})\in D_{0}$
s.t. for some $\lambda\in\Lambda_{S}$, $(\bar{s}_{1}(\lambda),\bar{e}_{1}(\lambda))=(\bar{s}_{2}(\lambda),\bar{e}_{2}(\lambda))$.
Take $\bar{s}_{1}\in\alpha_{1}\in\gamma$ and $\bar{s}_{2}\in\alpha_{2}\in\gamma$;
note that since $\bar{e}_{1}(\lambda)=\bar{e}_{2}(\lambda)$, $|\alpha_{1}[-\infty,\lambda]|=|\alpha_{2}[-\infty,\lambda]|$. 

Take $(\bar{s},\bar{e})\equiv(\bar{s}_{1}[-\infty,\lambda]\circ\bar{s}_{2}[\lambda,\infty],\bar{e}_{1}[-\infty,\lambda]\circ,\bar{e}_{2}[\lambda,\infty])$.
$\bar{s}\in|\alpha_{1}[-\infty,\lambda]|\circ\alpha_{2}[\lambda,\infty]=|\alpha_{2}[-\infty,\lambda]|\circ\alpha_{2}[\lambda,\infty]=\alpha_{2}$,
so $(\bar{s},\bar{e})\in D_{0}$ iff for all $\lambda^{\prime}\in\Lambda_{D}$,
$\bar{e}(\lambda^{\prime})=b_{\lambda^{\prime}}(|\alpha_{2}[-\infty,\lambda^{\prime}]|)$,
which follows from the fact that $\bar{s}\in\alpha_{2}$. -

B: $\mathbb{D}$ is a dynamic space

- For any $p\in\mathcal{P}_{D_{0}}=\mathcal{P}_{\mathbb{D}}$, $p$
is only realized at a single $\lambda\in\Lambda_{S}$ in $D_{0}$.
Because $\mathbb{D}$ is a union over ``shifted'' copies of $D_{0}$,
paths in $\mathbb{D}$ can only intersect if the belong to the same
copy; it follows that $\mathbb{D}$ is a dynamic space if $D_{0}$
is. -

C: $\mathbb{D}$ is homogeneous and $I$ is homogeneously realized

- Because for any $p\in\mathcal{P}_{\mathbb{D}}$, $p$ is realized
at a single $\lambda\in\Lambda_{S}$ in $D_{0}$, $D_{0}$ must be
homogeneous. From the nature of the construction of $\mathbb{D}$
from $D_{0}$ it's clear that $\mathbb{D}$ is also homogeneous, and
$I$ homogeneously realized in $D$. -

D: $I\!\rightarrow\:=I\!\rightarrow\!|\tilde{I}\Vert\!\rightarrow\!F\!\rightarrow$
and $\rightarrow\!F=\:\rightarrow\!I\!\rightarrow\!|\tilde{I}\Vert\!\rightarrow\!F$

- $D_{0}=\:\rightarrow\!I\!\rightarrow\!|\tilde{I}\Vert\!\rightarrow\!F\!\rightarrow$
and $I$ is only realized at $\lambda_{0}$, so for $D_{0}$, $I\!\rightarrow\:=I\!\rightarrow\!|\tilde{I}\Vert\!\rightarrow\!F\!\rightarrow$
and $\rightarrow\!F=\:\rightarrow I\!\rightarrow\!|\tilde{I}\Vert\!\rightarrow\!F$.
If this holds for $D_{0}$, it must also hold for $\mathbb{D}$ -

E: $Z$ is an environmental shell

- All elements of $Dom(Z)$ can be decomposed into system \& environmental
states, and $F$ has its own set of environmental states -

F: $Z$ is weakly unbiased

- Recall that $I=D_{0}(\lambda_{0})$. Choose any $e_{1}\in\mathcal{F}$,
and take $\alpha\in\gamma$, $\lambda_{1}+\lambda_{0}\in\Lambda_{S}$
s.t. $b_{\lambda_{0}+\lambda_{1}}(|\alpha[-\infty,\lambda_{0}+\lambda_{1}]|)=e_{1}$.
Take any $e\in\mathcal{E}\cdot Int(Z)$, $0\leq\lambda\leq\lambda_{1}$
s.t. $\Sigma_{e}\bigcap|\mathcal{O}_{e_{1}}[0,\lambda]|\neq\emptyset$.
It follows that $e=b_{\lambda_{0}+\lambda}(|\alpha[-\infty,\lambda_{0}+\lambda]|)$.
Therefore, if $e,e^{\prime}\in\mathcal{E}\cdot Int(Z)$ and for some
$\lambda\in\Lambda$, $\Sigma_{e}\bigcap|\mathcal{O}_{e_{1}}[0,\lambda]|\neq\emptyset$
and $\Sigma_{e^{\prime}}\bigcap|\mathcal{O}_{e_{1}}[0,\lambda]|\neq\emptyset$
then $e=e^{\prime}$. $Z$ must then be weakly unbiased -

G: $Z$ is all-reet

- For any $e\in\mathcal{F}$ there's only a single $\bar{e}[0,\lambda]\in\mathcal{E}\cdot\omega_{F}$,
so naturally $\Theta_{e}=(\mathcal{S}\cdot\Theta_{e})\times(\mathcal{E}\cdot\Theta_{e})$.

H: $Z$ is Boolean

- Since $Z$ is all-reet, Thm\ref{thm:When Boolean} can be used.

For any $\bar{s}[0,\lambda]\in\mathcal{O}_{F}$ there's only one $(\bar{s}[0,\lambda],\bar{e}[0,\lambda])\in\Theta_{F}$.
Since $\bar{e}(\lambda)\in\mathcal{F}$ there's only one $e\in\mathcal{F}$
s.t. $\bar{s}[0,\lambda]\in\mathcal{O}_{e}$. 

Take any $\bar{s}_{1}[0,\lambda_{1}],\bar{s}_{2}[0,\lambda_{2}]\in\mathcal{O}_{\mathcal{F}}$
s.t. $\lambda_{2}\geq\lambda_{1}$ and $\bar{s}_{1}[0,\lambda_{1}]=\bar{s}_{2}[0,\lambda_{1}]$.
For $\bar{p}_{1}[0,\lambda_{1}],\bar{p}_{2}[0,\lambda_{2}]\in\Theta_{F}$
s.t. $\mathcal{S}\cdot\bar{p}_{1}[0,\lambda_{1}]=\bar{s}_{1}[0,\lambda_{1}]$
and $\mathcal{S}\cdot\bar{p}_{2}[0,\lambda_{2}]=\bar{s}_{2}[0,\lambda_{2}]$,
$\bar{p}_{1}(\lambda_{1})=\bar{p}_{2}(\lambda_{1})\in F$, so $\lambda_{2}=\lambda_{1}$.
-

I: $Z$ properly decides $\gamma$

- Given any $\alpha\in\gamma$, there is one $\lambda_{\alpha}$ s.t.
$b_{\lambda_{\alpha}}(|\alpha[-\infty,\lambda_{\alpha}]|)\in\mathcal{F}$.
With $e=b_{\lambda_{\alpha}}(|\alpha[-\infty,\lambda_{\alpha}]|)$,
$\alpha=S_{\rightarrow\mathcal{O}_{e}^{\lambda_{0}}\rightarrow}$.
Thus, $Z$ decides $\gamma$. That $\gamma$ is properly decided follows
from the fact that $\gamma$ is an ideal partition. -
\end{proof}
The final theorem considers the special case in which the system is
described by a dynamic space:
\begin{thm}
If $\gamma$ is an ideal partition, and $S=\bigcup\gamma$ is a dynamic
space, then $\gamma$ is decided by a strongly unbiased ideal recorder.
\end{thm}

\begin{proof}
Use the same construction as the prior theorem. For every $(s,e)\in\mathcal{P}_{D_{0}}$,
there's an $\alpha\in\gamma$ and a $\lambda\in\Lambda_{S}$ s.t.
$e=b_{\lambda}(|\alpha[-\infty,\lambda]|)$ and $\mathcal{S}\cdot(D_{0})_{(s,e)\rightarrow}=(S_{|\alpha[-\infty,\lambda]|\rightarrow})_{(\lambda,s)\rightarrow}$.
If $S=\bigcup\gamma$ is a dynamic space, then for any $X\subset S$
s.t. $s\in X(\lambda)$ and $X=S_{X[-\infty,\lambda]\rightarrow}$,
$X_{(\lambda,s)\rightarrow}=S_{(\lambda,s)\rightarrow}.$ Therefore
$\mathcal{S}\cdot(D_{0})_{(s,e)\rightarrow}=S{}_{(\lambda,s)\rightarrow}$.
Since $\Sigma_{(s,e)\rightarrow}$ is simply $\mathcal{S}\cdot(D_{0})_{(s,e)\rightarrow}$
shifted by $\lambda_{0}$ and truncated at the $\lambda_{\alpha}$'s
(for definitions of $\lambda_{0}$ and $\lambda_{\alpha}$, see (3)
and (4) in the proof of the prior theorem), it follows immediately
that the recorder is strongly unbiased.
\end{proof}
These results culminate our formal investigation into experiments.
We conclude by further analyzing the make-up of ideal partitions.

\subsubsection{\label{subsec:ip composition}The Composition of Ideal Partitions}
\begin{rem*}
In continuation of prior practices, we will represent $S_{\alpha[-\infty,\lambda]\rightarrow}$
by $\alpha[-\infty,\lambda]\!\rightarrow$ in cases where $S$ is
understood.
\end{rem*}
Given the significance of ideal partitions, it will be helpful to
understand more about their structure, particularly how they unfold
over time. To do so, we start with the simplest partitions, those
whose outcomes are differentiated in no time at all:
\begin{defn}
For dynamic set $S$

$\alpha\subset S$ is a \emph{forward moment at $\lambda$} if $\alpha=\bigcup_{\epsilon>0}S_{\rightarrow\alpha[\lambda,\lambda+\epsilon]\rightarrow}$.

$\alpha\subset S$ is a \emph{backward moment at $\lambda$} if $\alpha=\bigcup_{\epsilon>0}S_{\rightarrow\alpha[\lambda-\epsilon,\lambda]\rightarrow}$.

A partition of $S$, $\gamma$, is a \emph{forward/backward moment
partition at $\lambda$} if it is composed entirely of forward/backward
moments at $\lambda$.
\end{defn}

Moments are measurements that are completed in an instant. Such measurements
are commonly considered in calculus. For example, moment measurements
on classical particles include measurements of position and/or velocity
and/or acceleration, etc. 

Moment partitions have a useful feature that many ideal partitions
lack:
\begin{thm}
\label{thm:Moments closed under unions}If $\gamma$ is a forward/backward
moment partition at $\lambda$, and $A\subset\gamma$, then $(\gamma\setminus A)\bigcup\{\bigcup A\}$
is a forward/backward moment partition at $\lambda$
\end{thm}

\begin{proof}
Immediate from the definition of moment partitions
\end{proof}
It follows that if the measurements corresponding to $\gamma$ and
$\gamma^{\prime}=(\gamma\setminus A)\bigcup\{\bigcup A\}$ can both
be performed, and the probability of an outcome occurring is independent
of the set of outcomes as a whole, then $P(A)=P(\{\bigcup A\})$.
Statements similar to Thm \ref{thm:Moments closed under unions} do
not hold in general for ideal partitions, which makes non-additive
probabilities for extended time experiments unsurprising. Note that
in quantum theory, moment measurements such as individual position
or momentum measurements, do have additive probabilities, while extended
time measurements generally do not.

The major result of this section will be that ideal partitions are
branching sequences of moment partitions. To obtain this result, we
first show that branching sequences of moment partitions are ideal
partitions, and then show that any ideal partition can be decomposed
into such sequences. 

To compose a branching sequence of moment partitions, start with a
partition containing a single element, where that element is a dynamic
space. This corresponds to the case where no measurements are performed
at all; such a ``partition'' is ideal. Then sequentially append
moment partitions by selecting an outcome from the current partition,
splitting it by applying a moment partition at any time after the
outcome has been obtained, and (optionally) repeating this operation
on an element of the new partition. Because measurements can alter
system dynamics, when appending moments to outcomes we need to take
such changes into account. Any change to the dynamics is allowed,
so long as it satisfies some minor constraints, such as not altering
system dynamics prior to the measurement, and only effecting outcomes
on which the measurement takes place. This is captured in the following
definition:
\begin{defn}
\label{def:Adding moment}If $\gamma$ is an ideal partition, $S\equiv\bigcup\gamma$,
and $\alpha\in\gamma$, a moment partition can be \emph{appended to
$\alpha$ at $\lambda$} if $\alpha=S_{\alpha[-\infty,\lambda]\rightarrow}$,
or if the moment partition is forward and for all $\lambda^{\prime}>\lambda$,
$\alpha=S_{\alpha[-\infty,\lambda^{\prime}]\rightarrow}$. (The second
condition allows us to append a forward partition at $\lambda$, even
if outcome $\alpha$ ended with a forward partition at $\lambda$.
The prior condition, that $\alpha=S_{\alpha[-\infty,\lambda]\rightarrow}$,
allows this for backward moments.) 

$\gamma^{\prime}$ is the result of appending a moment partition to
$\alpha$ at $\lambda$ if for some $\gamma_{m}\subset\gamma^{\prime}$:

1)\emph{ }$\gamma^{\prime}\setminus\gamma_{m}=\gamma\setminus\{\alpha\}$

2) For any $\lambda^{\prime}<\lambda$, $(\bigcup\gamma_{m})[-\infty,\lambda^{\prime}]=\alpha[-\infty,\lambda^{\prime}]$

3) $\gamma_{m}$ is a moment partition of $\bigcup\gamma_{m}$ at
$\lambda$ 

4) For all $\lambda^{\prime}<\lambda$, $\bigcup\gamma_{m}$ is unbiased
with respect to $(\lambda^{\prime},\gamma_{m})$
\end{defn}

Traditional quantum theory provides one of the most dramatic examples
of system dynamics being altered by moment measurements. There, the
entire set of available states may change based on the measurement
being performed. However, even under those circumstances, the above
constraints are still respected.

We can now show that appending a moment partition to an ideal partition
(including the trivial ideal partition consisting of a single dynamic
space), results is a new ideal partition.
\begin{thm}
\label{thm:moments are ips}If $\gamma$ is an ideal partition, $\alpha\in\gamma$,
and $\gamma^{\prime}$ is the result of appending a moment partition
to $\alpha$, then $\gamma^{\prime}$ is an ideal partition:
\end{thm}

\begin{proof}
We only need to show that for all $\beta\in\gamma^{\prime}$, $\lambda\in\Lambda_{\bigcup\gamma^{\prime}}$,
$\beta=|\beta[-\infty,\lambda]|_{\gamma^{\prime}}\circ\beta[\lambda,\infty]$,
as the other aspects of ideal partitions clearly hold.

Define $\gamma_{m}\equiv\gamma^{\prime}\setminus\gamma$.

For all $\beta\in\gamma^{\prime}\setminus\gamma_{m}$, all $\lambda\in\Lambda_{\bigcup\gamma^{\prime}}$,
$\beta[-\infty,\lambda]\subset|\beta[-\infty,\lambda]|_{\gamma^{\prime}}\subset|\beta[-\infty,\lambda]|_{\gamma}$.
Therefore $\beta\subset|\beta[-\infty,\lambda]|_{\gamma^{\prime}}\circ\beta[\lambda,\infty]\subset|\beta[-\infty,\lambda]|_{\gamma}\circ\beta[\lambda,\infty]$.
Because $\beta=|\beta[-\infty,\lambda]|_{\gamma}\circ\beta[\lambda,\infty]$,
$\beta=|\beta[-\infty,\lambda]|_{\gamma^{\prime}}\circ\beta[\lambda,\infty]$. 

Assume the moment partition is appended to $\alpha$ at, $\lambda_{m}$.

For $\beta\in\gamma_{m}$, $\lambda<\lambda_{m}$: By the nature of
an ideal partition, $(|\alpha[-\infty,\lambda]|_{\gamma})_{\rightarrow(\lambda,\alpha(\lambda))}=\alpha_{\rightarrow(\lambda,\alpha(\lambda))}$.
Since $\beta(\lambda)\subset\alpha(\lambda)$, $(|\alpha[-\infty,\lambda]|_{\gamma})_{\rightarrow(\lambda,\beta(\lambda))}=\alpha_{\rightarrow(\lambda,\beta(\lambda))}$
so by (2) in the above definition, $(|\alpha[-\infty,\lambda]|_{\gamma})_{\rightarrow(\lambda,\beta(\lambda))}=(\bigcup\gamma_{m})_{\rightarrow(\lambda,\beta(\lambda))}$.
By (4), $\bigcup\gamma_{m}$ is unbiased with respect to $(\lambda,\gamma_{m})$,
so $|\alpha[-\infty,\lambda]|_{\gamma}\circ\beta[\lambda,\infty]\subset\bigcup\gamma_{m}$,
and so by (3), $\beta=|\alpha[-\infty,\lambda]|_{\gamma}\circ\beta[\lambda,\infty]$.
Since $|\beta[-\infty,\lambda]|_{\gamma^{\prime}}\subset|\alpha[-\infty,\lambda]|_{\gamma}$,
$\beta\subset|\beta[-\infty,\lambda]|_{\gamma^{\prime}}\circ\beta[\lambda,\infty]\subset|\alpha[-\infty,\lambda]|_{\gamma}\circ\beta[\lambda,\infty]$,
and so $\beta=|\beta[-\infty,\lambda]|_{\gamma^{\prime}}\circ\beta[\lambda,\infty]$. 

For $\beta\in\gamma_{m}$, $\lambda\geq\lambda_{m}$: By the restriction
on when a moment partition can be appended to an outcome, for $\lambda\geq\lambda_{m}$,
$|\beta[-\infty,\lambda]|_{\gamma^{\prime}}=|\beta[-\infty,\lambda]|_{\gamma_{m}}$.
By (4) in the above definition, $|\beta[-\infty,\lambda]|_{\gamma_{m}}\circ\beta[\lambda,\infty]\subset\bigcup\gamma_{m}$.
By (3) and the nature of a moment partition, for any $\bar{s}\in|\beta[-\infty,\lambda]|\circ\beta[\lambda,\infty]$,
$\bar{s}\in\beta$.
\end{proof}
Thm \ref{thm:moments are ips} means that branching sequences of moment
partitions form ideal partitions. Strictly speaking, we have only
shown this holds for finite sequences, but it is reasonable to conclude
that it holds for any cardinality, so long as each of the sequences
complete at a finite time. 

We now seek to show the converse, that all ideal partitions are branching
sequences of moment partitions. Given that outcomes of ideal partitions
always satisfy $\alpha=|\alpha[-\infty,\lambda]|\circ\alpha[\lambda,\infty]|$,
ideal partitions, and the recorders that generate them, are characterized
by how $|\alpha[-\infty,\lambda]|\!\!\!\rightarrow$ sheds paths as
time progresses. This shedding is due to the environment accumulating
information about the system. More general types of recorders are
not wholly characterized by this process: If a recorder is not Boolean
then $|\alpha[-\infty,\lambda]||\!\!\rightarrow$ may not evolve to
$\alpha$, while if a recorder is not all-reet, then the recorder
is characterized both by how information is gained, and how some of
this information is later lost. 

To understand how ideal partitions unfold with time, we start with
the following definition:
\begin{defn}
If $\gamma$ is an ideal partition, $S=\bigcup\gamma$, and $\alpha\in\gamma$
then

$\lnot\alpha{}_{(\lambda,f);\gamma}\equiv\{\bar{s}\in S:\bar{s}[-\infty,\lambda]\in|\alpha[-\infty,\lambda]|\:and\:for\:all\:\lambda^{\prime}>\lambda,\:\bar{s}[-\infty,\lambda^{\prime}]\notin|\alpha[-\infty,\lambda^{\prime}]|\}$

$\lnot\alpha{}_{(\lambda,b);\gamma}\equiv\{\bar{s}\in S:\bar{s}[-\infty,\lambda]\notin|\alpha[-\infty,\lambda]|\:and\:for\:all\:\lambda^{\prime}<\lambda,\:\bar{s}[-\infty,\lambda^{\prime}]\in|\alpha[-\infty,\lambda^{\prime}]|\}$
\end{defn}

$\lnot\alpha{}_{(\lambda,b);\gamma}$ is the collection of paths that
break off from $|\alpha[-\infty,\lambda^{\prime}]|\!\rightarrow$
at $\lambda^{\prime}=\lambda$, while $\lnot\alpha{}_{(\lambda,f);\gamma}$
is the collection of paths that break off from $|\alpha[-\infty,\lambda^{\prime}]|\!\rightarrow$
just after $\lambda^{\prime}=\lambda$. The ``b'' subscript is for
``backward'', and the ``f'' is for ``forward''.

To relate these sets to the outcomes in $\gamma$, we use the following
theorem:
\begin{thm}
\label{thm:ip lemma}If $\gamma$ is an ideal partition and $\alpha\in\gamma$: 

1) If $\bar{p}\in\lnot\alpha{}_{(\lambda,b);\gamma}$, and $\bar{p}\in\beta\in\gamma$,
then $|\mathbf{\beta}[-\infty,\lambda]|\!\rightarrow\,\subset\lnot\alpha{}_{(\lambda,b);\gamma}$

2) If $\bar{p}\in\lnot\alpha{}_{(\lambda,f);\gamma}$, and $\bar{p}\in\beta\in\gamma$,
then $\bigcap_{\lambda^{\prime}>\lambda}|\beta[-\infty,\lambda^{\prime}]|\!\rightarrow\,\subset\lnot\alpha{}_{(\lambda,f);\gamma}$
\end{thm}

\begin{proof}
For all $\lambda^{\prime}$, either $|\beta[-\infty,\lambda^{\prime}]|=|\alpha[-\infty,\lambda^{\prime}]|$
or $|\beta[-\infty,\lambda^{\prime}]|\bigcap|\alpha[-\infty,\lambda^{\prime}]|=\emptyset$,
so if $\bar{p}[-\infty,\lambda^{\prime}]\in|\alpha[-\infty,\lambda^{\prime}]$|
then $|\beta[-\infty,\lambda^{\prime}]|=|\alpha[-\infty,\lambda^{\prime}]|$,
and if $\bar{p}[-\infty,\lambda^{\prime}]\notin|\alpha[-\infty,\lambda^{\prime}]|$
then $|\beta[-\infty,\lambda^{\prime}]|\bigcap|\alpha[-\infty,\lambda]|=\emptyset$.
\end{proof}
This allows us to define moment partitions on the various $|\alpha[-\infty,\lambda]|\!\rightarrow$
based on when they shed their paths:
\begin{defn}
\label{def:ip moments}If $\gamma$ is an ideal partition, $S=\bigcup\gamma$,
and $\alpha\in\gamma$ then

$[\alpha]{}_{(\lambda,b);\gamma}\equiv\{|\beta[-\infty,\lambda]|\!\rightarrow\,:\beta\in\gamma\:and\:either\:\beta=\alpha\:or\:\beta\subset\lnot\alpha{}_{(\lambda,b);\gamma}\}$ 

$[\alpha]{}_{(\lambda,f);\gamma}\equiv\{\bigcap_{\lambda^{\prime}>\lambda}|\beta[-\infty,\lambda^{\prime}]|\!\rightarrow\,:\beta\in\gamma\:and\:either\:\beta=\alpha\:or\:\beta\subset\lnot\alpha{}_{(\lambda,f);\gamma}\}$ 
\end{defn}

We can now prove the heart of the desired result:
\begin{thm}
\label{thm:ip composition}If $\gamma$ is an ideal partition, and
$\alpha\in\gamma$, then $[\alpha]{}_{(\lambda,f);\gamma}$ is a forward
moment partition of $|\alpha[-\infty,\lambda]|\!\!\rightarrow$ at
$\lambda$, and $[\alpha]{}_{(\lambda,b);\gamma}$ is a backward moment
partition of $\bigcap_{\lambda^{\prime}<\lambda}|\alpha[-\infty,\lambda^{\prime}]|\!\rightarrow$
at $\lambda$
\end{thm}

\begin{proof}
It is clear that $[\alpha]{}_{(\lambda,f);\gamma}$ is a partition
of $|\alpha[-\infty,\lambda]|\!\rightarrow$, and $[\alpha]{}_{(\lambda,b);\gamma}$
is a partition of $\bigcap_{\lambda^{\prime}<\lambda}|\alpha[-\infty,\lambda^{\prime}]|\!\rightarrow$.
It remains to show that they are moment partitions.

Let's start with $[\alpha]{}_{(\lambda,b);\gamma}$. For any $\bar{p}\in|\beta[-\infty,\lambda]|\!\rightarrow\,\in[\alpha]{}_{(\lambda,b);\gamma}$,
take any $\bar{p}^{\prime}\in\bigcap_{\lambda^{\prime}<\lambda}|\alpha[-\infty,\lambda^{\prime}]|\!\rightarrow$
s.t. for some $\epsilon>0$, $\bar{p}^{\prime}[\lambda-\epsilon,\lambda]=\bar{p}[\lambda-\epsilon,\lambda]$.
Let's say $\bar{p}\in\delta\in\gamma$ and $\bar{p}^{\prime}\in\eta\in\gamma$.
Note that $|\delta[-\infty,\lambda]|=|\beta[-\infty,\lambda]|$. Since
$\bar{p}^{\prime}\in\bigcap_{\lambda^{\prime}<\lambda}|\alpha[-\infty,\lambda^{\prime}]|\!\rightarrow$,
and by the nature of $[\alpha]{}_{(\lambda,b);\gamma}$, $|\eta[-\infty,\lambda-\epsilon]|=|\delta[-\infty,\lambda-\epsilon]|=|\alpha[-\infty,\lambda-\epsilon]|$.
Since $\bar{p}^{\prime}[-\infty,\lambda-\epsilon]\in|\delta[-\infty,\lambda-\epsilon]|$
and $\bar{p}^{\prime}[\lambda-\epsilon,\lambda]=\delta[\lambda-\epsilon,\lambda]$,
$\bar{p}^{\prime}[-\infty,\lambda]\in|\delta[-\infty,\lambda-\epsilon]|\circ\delta[\lambda-\epsilon,\lambda]=\delta[-\infty,\lambda]$.
Therefore $\bar{p}^{\prime}\in|\delta[-\infty,\lambda]|\!\rightarrow\:=|\beta[-\infty,\lambda]|\!\rightarrow$.

The forward case is nearly identical. For any $\bar{p}\in\bigcap_{\lambda^{\prime}>\lambda}|\beta[-\infty,\lambda^{\prime}]|\!\rightarrow\,\in[\alpha]{}_{(\lambda,f);\gamma}$,
take any $\bar{p}^{\prime}\in|\alpha[-\infty,\lambda]|\!\rightarrow$
s.t. for some $\epsilon>0$, $\bar{p}^{\prime}[\lambda,\lambda+\epsilon]=\bar{p}[\lambda,\lambda+\epsilon]$.
Let's say $\bar{p}\in\delta\in\gamma$ and $\bar{p}^{\prime}\in\eta\in\gamma$.
Note that $\bigcap_{\lambda^{\prime}>\lambda}|\delta[-\infty,\lambda^{\prime}]|\!\rightarrow\:=\bigcap_{\lambda^{\prime}>\lambda}|\beta[-\infty,\lambda^{\prime}]|\!\rightarrow$.
Since $\bar{p}^{\prime}[-\infty,\lambda]\in|\alpha[-\infty,\lambda]|=|\delta[-\infty,\lambda]|$,
and $\bar{p}^{\prime}[\lambda,\lambda+\epsilon]\in\delta[\lambda,\lambda+\epsilon]$,
$\bar{p}^{\prime}[-\infty,\lambda+\epsilon]\in|\delta[-\infty,\lambda]|\circ\delta[\lambda,\lambda+\epsilon]=\delta[-\infty,\lambda+\epsilon]$.
Therefore $\bar{p}^{\prime}\in\bigcap_{\lambda^{\prime}>\lambda}|\delta[-\infty,\lambda^{\prime}]|\!\rightarrow\:=\bigcap_{\lambda^{\prime}>\lambda}|\beta[-\infty,\lambda^{\prime}]|\!\rightarrow$.
\end{proof}
To complete the picture, choose any $\lambda_{0}$ such that for all
$\beta\in\gamma$, $\beta=S[-\infty,\lambda_{0}]\circ\beta[\lambda_{0},\infty]$.
The set $\{|\beta[-\infty,\lambda_{0}]|\!\rightarrow\,:\beta\in\gamma\}$
is a moment partition, as can be seen from the fact that, for all
$\beta\in\gamma$, $|\beta[-\infty,\lambda_{0}]|\!\rightarrow\:=\:\rightarrow\!(\lambda_{0},|\beta[-\infty,\lambda_{0}]|(\lambda_{0}))\!\rightarrow$.
This is the starting measurement of the ideal partition (possibly
containing only a single outcome). From there, the outcomes split
by applying moment partitions of the form $[\alpha]{}_{(\lambda,b);\gamma}$
and $[\alpha]{}_{(\lambda,f);\gamma}$. Thus, an ideal partition unfolds
as a branching tree, each node of the tree being a moment partition
that acts to split outcomes from each other. 

Together, Thms \ref{thm:moments are ips} and \ref{thm:ip composition}
mean that ideal partitions are precisely the partitions that unfold
in this manner, with the mild additional restriction that the branching
must start an end at finite times.

\section{\label{part:Probabilities}Experimental probabilities}

\subsection{Initial considerations}

In this part we will turn our attention from experiments to experimental
probabilities. In order to be able to reliably assign experimental
probabilities, some of the experimental axioms will have to be extended
to hold probabilistically:
\begin{itemize}
\item $D$ is homogeneous: If a closed experimental system is in any set
$p$, the probabilities of what will happen, or what has happened,
are independent of time.
\item The unbiased condition: For a properly designed experiment, the system's
environment do not unduly effect the likelihoods of the experimental
outcomes.
\item The experimental start state: Beyond starting in $I$, we also assume
that the probability distribution of the system's initial state is
the same for all experimental runs.
\end{itemize}
These will be sufficient to ensure that experimental probabilities
can be reliably determined.

\subsection{\label{subsec:Dynamic-Probability-Spaces}Dynamic Probability Spaces}

We start by reviewing probabilities for individual experiments and
collections of experiments. It will be seen that individual experiments
conform to traditional probability theory, while collections of experiments
may not.

For an individual experiment, an outcome's probability is the expected
ratio of the number of times the outcome obtains, to the total number
of times the experiment is run; the probability for a set of the experiment's
outcomes is assigned similarly. This is sufficient to motivate the
laws of experimental probability: All probabilities have values between
0 and 1, if $\gamma$ is the collection of all outcomes for an experiment
then $P(\gamma)=1$, and if $A$ and $B$ are disjoint collections
of outcomes from the same experiment then $P(A\bigcup B)=P(A)+P(B)$.

These rules are easily formalized. In keeping with our earlier results
we assume that, for experiments of interest, the complete set of outcomes
form an ideal partition. We then define an \emph{experimental probability
space} (eps) to be a triple, $(\gamma,\Sigma,P)$, where $\gamma$
is an ideal partition, $\Sigma$ is a set of subsets of $\gamma$,
and $P$ is a function $P:\Sigma\rightarrow[0,1]$, which satisfy:

1) $\gamma\in\Sigma$

2) If $\sigma\in\Sigma$ then $\gamma\setminus\sigma\in\Sigma$

3) If $\sigma_{1},\sigma_{2}\in\Sigma$ then $\sigma_{1}\bigcup\sigma_{2}\in\Sigma$

4) $P(\gamma)=1$

5) If $\sigma_{1},\sigma_{2}\in\Sigma$ and $\sigma_{1}\bigcap\sigma_{2}=\emptyset$
then $P(\sigma_{1}\bigcup\sigma_{2})=P(\sigma_{1})+P(\sigma_{2})$

These are, of course, the traditional rules governing probabilities.
In axiomatic probability theory, (3) and (5) are are often extended
to hold for countable subsets of $\varSigma$; that variation will
be considered in Section \ref{subsec:Convergence}. 

Conditions (2) \& (3) are equivalent to the single statement: If $\sigma_{1},\sigma_{2}\in\Sigma$
then $\sigma_{1}\setminus\sigma_{2}\in\Sigma$. This allows us to
replace the 5 assumptions enumerated above with:

1) $\gamma\in\Sigma$

2) If $\sigma_{1},\sigma_{2}\in\Sigma$ then $\sigma_{1}\setminus\sigma_{2}\in\Sigma$

3) $P(\gamma)=1$

4) If $\sigma_{1},\sigma_{2}\in\Sigma$ and $\sigma_{1}\bigcap\sigma_{2}=\emptyset$
then $P(\sigma_{1}\bigcup\sigma_{2})=P(\sigma_{1})+P(\sigma_{2})$

To understand a physical system, it is rarely sufficient to only have
to consider a single experiment. Because many different sorts of experiments
can be performed on a physical system, more often than not we are
interested in collections of eps's. Such collections form a coherent
structure when their probabilities are consistent with each other.
To be precise, we say that a pair of eps's, $(\gamma_{1},\Sigma_{1},P_{1})$
and $(\gamma_{2},\Sigma_{2},P_{2})$, are \emph{consistent} if:

1) $\gamma_{1}\bigcap\gamma_{2}\in\Sigma_{1}$ and $\gamma_{1}\bigcap\gamma_{2}\in\Sigma_{2}$

2) For any $\sigma\subset\gamma_{1}\bigcap\gamma_{2}$, $\sigma\in\Sigma_{1}$
iff $\sigma\in\Sigma_{2}$

3) For any $\sigma\in\Sigma_{1}\bigcap\Sigma_{2}$, $P_{1}(\sigma)=P_{2}(\sigma)$

The second and third consistency conditions require that, on the set
of common outcomes, the two probability spaces are identical. The
first condition ensures that the two experiments are comparable on
the set of common outcomes; without it, we could have a $\sigma\in\Sigma_{1}$
that overlaps $\gamma_{1}\bigcap\gamma_{2}$, but does not contain
any non-empty subsets that are elements of $\Sigma_{2}$. 

Consistency can be extended to collections of experiments by saying
that a collection of eps's, $X$, is consistent if, for any $x,y\in X$,
$x$ and $y$ are consistent. Consistent collections of eps's can
be conveniently represented as \emph{dynamic probability space}:
\begin{defn}
A \emph{dynamic probability space} (dps) is a triple, $(X,T,P)$ where
$X$ and $T$ are sets, $P:T\rightarrow[0,1]$, and these satisfy.:

1) If $\gamma\in X$ then $\gamma$ is an ideal partition

2) If $t\in T$ then for some $\gamma\in X$, $t\subset\gamma$

3) $X\subset T$

4) If $t_{1},t_{2}\in T$ then $t_{1}\setminus t_{2}\in T$

5) If $\gamma\in X$ then $P(\gamma)=1$

6) If $t_{1},t_{2}\in T$ are disjoint, and $t_{1}\bigcup t_{2}\in T$,
then $P(t_{1}\bigcup t_{2})=P(t_{1})+P(t_{2})$
\end{defn}

All dps's are equivalent to some consistent sets of eps's, and vice
versa. To extract the eps's from a dps, iterate through the $\gamma\in X$,
and define $\Sigma_{\gamma}\equiv\{t\in T:t\subset\gamma\}$ and $P_{\gamma}\equiv P|_{\Sigma_{\gamma}}$.
The resulting $(\gamma,\Sigma_{\gamma},P_{\gamma})$'s form a consistent
set of eps's that contains all the information present in the original
dps. Going the other way, starting with any consistent set of eps's,
$Y$, define $X_{Y}\equiv\{\gamma:(\gamma,\Sigma,P)\in Y\}$, $T_{Y}\equiv\bigcup_{(\gamma,\Sigma,P)\in Y}\Sigma$,
and $P_{Y}:T_{Y}\rightarrow[0,1]$ s.t. if $(\gamma,\Sigma,P)\in Y$
and $t\in\Sigma$ then $P_{Y}(t)=P(t)$. $(X_{Y},T_{Y},P_{Y})$ is
a dps that contains all the information present in the original collection
of eps's.

It's important to stress that, if $(X,T,P)$ is a dps, and $\{\alpha\},\{\beta\},\{\alpha\bigcup\beta\}\in T$,
then condition (6) demands that $P(\{\alpha,\beta\})=P(\{\alpha\})+P(\{\beta\})$;
it does not mean that $P(\{\alpha,\beta\})=P(\{\alpha\bigcup\beta\})$.
In quantum systems it is always the case that $P(\{\alpha,\beta\})=P(\{\alpha\})+P(\{\beta\})$,
though often the case that $P(\{\alpha\bigcup\beta\})\neq P(\{\alpha\})+P(\{\beta\})$.
The later is why quantum probabilities are referred to as ``non-additive''.
Also note that, unlike traditional probability spaces, arbitrary finite
unions of elements of $T$ cannot be expected to be elements of $T$.
Arbitrary finite intersections, on the other hand, are elements of
$T$.

We next will expand dps's into a larger structure that renders their
content more accessible.

\subsection{\label{sec:Generalized-Probability-Spaces}Generalized Probability
Spaces and the $(X_{\mathbb{\omega}},T_{\mathbb{\omega}},P_{\omega})$
construction}

Given any dps, $(X,T,P)$, useful probabilistic information can be
extracted from $X$ and $T$ alone. Preparatory to doing so, it will
be helpful to generalize the definition of dps's by removing the explicit
reference to ideal partitions:
\begin{defn}
An \emph{event-algebra} is a double, $(X,T)$, where $X$ and $T$
are collection of sets that satisfy:

1) If $t\in T$ then for some $\gamma\in X$, $t\subset\gamma$

2) If $\gamma\in X$ then $\gamma\in T$

3) If $t_{1},t_{2}\in T$ then $t_{1}\setminus t_{2}\in T$

A \emph{generalized probability space} (gps) is a triple, $(X,T,P)$,
where $(X,T)$ is an event-algebra and $P:T\rightarrow[0,1]$ s.t.

1) If $\gamma\in X$ then $P(\gamma)=1$

2) If $t_{1},t_{2}\in T$ are disjoint and $t_{1}\bigcup t_{2}\in T$
then $P(t_{1}\bigcup t_{2})=P(t_{1})+P(t_{2})$.
\end{defn}

When $X$ contains only a single element, this is equivalent to traditional
probability theory. When $X$ consists of ideal partitions, the gps
is a dps.

For an event algebra, a particularly useful operation is the ``not''
operation, which gives the collection of complements to an element
of $T$:
\begin{defn}
For event-algebra $(X,T)$, $t\in T$, $\neg t\equiv\{t^{\prime}\in T:t^{\prime}\bigcap t=\emptyset\:\&\:t^{\prime}\bigcup t\in X\}$.

For $A\subset T$, $\neg A\equiv\bigcup_{t\in A}\neg t$.
\end{defn}

We can compound $\neg$'s to form $\neg^{(n)}$ for any $n$ by defining
$\neg^{(1)}A\equiv\neg A$ and $\neg^{(n+1)}A\equiv\neg(\neg^{(n)}A)$.
For any gps on an event-algebra, if $t^{\prime}\in\neg^{(n)}t$, and
$n$ is even, then $P(t^{\prime})=P(t)$, while if $n$ is odd, then
$P(t^{\prime})+P(t)=1$. This motivates the following equivalence
class:
\begin{defn}
If $(X,T)$ is an event-algebra and $t\in T$ then $[t]\equiv\bigcup_{n\in\mathbb{N}^{+}}\neg^{(2n)}t$.
\end{defn}

It's easy to see that $[t]$ is an equivalence class, and that $\neg[t]=\bigcup_{n\in\mathbb{N}^{+}}\neg^{(2n+1)}t$.
These have the above mentioned property that if $t^{\prime}\in[t]$
then $P(t^{\prime})=P(t)$, and if $t^{\prime}\in\neg[t]$, $P(t^{\prime})+P(t)=1$.
Thus $[t]$ and $\neg[t]$ yield information that may not be explicitly
present in $(X,T,P)$.

We can obtain still further information by using $\neg[t]$ to create
a more expansive gps. To do so, start by defining the ``$\Join$''
relation by: If $t_{1}\in\neg[t_{2}]$, $t\subset t_{1}$, $t^{\prime}\subset t_{2}$,
and $t,t^{\prime}\in T$ then $t\Join t^{\prime}$. 
\begin{thm}
\label{thm:Feasibility of P1}If $(X,T,P)$ is a gps, $t_{1}\Join t_{2}$,
$t_{3}\Join t_{4}$, and $t_{1}\bigcup t_{2}=t_{3}\bigcup t_{4}$,
then $P(t_{1})+P(t_{2})=P(t_{3})+P(t_{4})\in[0,1]$.
\end{thm}

\begin{proof}
Take $t_{1}^{\prime}=t_{1}\bigcap t_{3}$, $t_{2}^{\prime}=t_{1}\bigcap t_{4}$,
$t_{3}^{\prime}=t_{2}\bigcap t_{3}$, $t_{4}^{\prime}=t_{2}\bigcap t_{4}$.
All $t_{i}^{\prime}$ are disjoint, all are elements of $T$, and
each $t_{i}$ is a union of two of the $t_{j}^{\prime}$, so $P(t_{1})+P(t_{2})=P(t_{1}^{\prime})+P(t_{2}^{\prime})+P(t_{3}^{\prime})+P(t_{4}^{\prime})=P(t_{3})+P(t_{4})$.

Since $t_{1}\Join t_{2}$ there's $t_{5}\in T$, $t_{6}\in\neg[t_{5}]$
s.t. $t_{1}\subset t_{5}$, and $t_{2}\subset t_{6}$. Since $t_{6}\in\neg[t_{5}]$,
$P(t_{5})+P(t_{6})=1$, and so $P(t_{1})+P(t_{2})\in[0,1]$.
\end{proof}
$\neg[t]$ and $t_{1}\Join t_{2}$ can be used to construct a new,
larger gps, $(X_{1},T_{1},P_{1})$, as follows. First, define $X_{1}$
by: For any $t_{1}\in T$, $t_{2}\in\neg[t_{1}]$, $t_{1}\bigcup t_{2}\in X_{1}$.
It's clear that $X$ is a subset of $X_{1}$. Next construct $T_{1}:$
For any $t_{1},t_{2}\in T$ s.t. $t_{1}\Join t_{2}$, $t_{1}\bigcup t_{2}\in T_{1}$.
Since $\emptyset=t\setminus t\in T$, and for any $t\in T$, $t\Join\emptyset$,
$T$ is a subset of $T_{1}$. Finally, we use the above theorem to
define function $P_{1}$: For any $t\in T_{1}$, take any $t_{1},t_{2}\in T$
s.t. $t_{1}\Join t_{2}$ and $t_{1}\bigcup t_{2}=t$, and assign $P_{1}(t)=P(t_{1})+P(t_{2})$. 
\begin{thm}
\label{thm:T1 is a gps}1) If $(X,T)$ is an event-algebra then $(X_{1},T_{1})$
is an event-algebra

2) If $(X,T,P)$ is a gps then $(X_{1},T_{1},P_{1})$ is a gps.
\end{thm}

\begin{proof}
1) Event-algebra axioms (1) and (2) clearly hold.

For axiom (3), given any $t_{1},t_{2}\in T_{1}$, there exist $t_{1}^{\prime},t_{1}^{\prime\prime},t_{2}^{\prime},t_{2}^{\prime\prime}\in T$
s.t. $t_{i}^{\prime}\Join t_{i}^{\prime\prime}$ and $t_{i}=t_{i}^{\prime}\bigcup t_{i}^{\prime\prime}$.
Define $t_{3}\equiv(t_{1}^{\prime}\setminus t_{2}^{\prime})\setminus t_{2}^{\prime\prime}$
and $t_{4}\equiv(t_{1}^{\prime\prime}\setminus t_{2}^{\prime})\setminus t_{2}^{\prime\prime}$.
$t_{3},t_{4}\in T$. Since $t_{3}\subset t_{1}^{\prime}$ and $t_{4}\subset t_{1}^{\prime\prime}$,
$t_{3}\Join t_{4}$, so $t_{3}\bigcup t_{4}\in T_{1}$. $t_{1}\setminus t_{2}=t_{3}\bigcup t_{4}$
so $t_{1}\setminus t_{2}\in T_{1}$.

2) Follows from (1) and Thm \ref{thm:Feasibility of P1}.
\end{proof}
Note that if $(X,T,P)$ is a dps then $(X_{1},T_{1},P_{1})$ may not
be a dps, because $X_{1}$ may not be composed entirely of ideal partitions.
This is why gps's were introduced. 

We can now define $(X_{2},T_{2},P_{2})\equiv((X_{1})_{1},(T_{1})_{1},(P_{1})_{1})$;
which is to say, $(X_{2},T_{2},P_{2})$ is formed from $(X_{1},T_{1},P_{1})$
in precisely the same way as $(X_{1},T_{1},P_{1})$ was formed from
$(X,T,P)$. As $(X_{1},T_{1},P_{1})$ is a gps, $(X_{2},T_{2},P_{2})$
must also be a gps. Continuing in this vein we can construct $(X_{3},T_{3},P_{3})$,
$(X_{4},T_{4},P_{4})$, etc. Finally, we can define $X_{\mathbb{\omega}}\equiv\bigcup_{n\in\mathbb{N}^{+}}X_{n}$,
$T_{\mathbb{\omega}}\equiv\bigcup_{n\in\mathbb{N}^{+}}T_{n}$, and
$P_{\omega}:T_{\mathbb{\omega}}\rightarrow[0,1]$ s.t. if $t\in T_{n}$
then $P_{\mathbb{N}}(t)=P_{n}(t)$. $(X_{\mathbb{\omega}},T_{\mathbb{\omega}},P_{\omega})$
is a gps (proved below); indeed, it is a \emph{simple} gps:
\begin{defn}
A gps, $(X,T,P)$ is \emph{simple} if for all $t\in T$, $\neg t=\neg[t]$
\end{defn}

It follows that a gps is simple if and only if $\neg t=\neg^{(3)}t$.
For a simple gps, $(X_{1},T_{1},P_{1})=(X,T,P)$, and so $(X_{\mathbb{\omega}},T_{\mathbb{\omega}},P_{\omega})=(X,T,P)$.
\begin{thm}
If $(X,T,P)$ is a gps then $(X_{\mathbb{\omega}},T_{\mathbb{\omega}},P_{\omega})$
is a simple gps
\end{thm}

\begin{proof}
That $(X_{\mathbb{\omega}},T_{\mathbb{\omega}})$ is an event-algebra
follows from the fact that if $t\in T_{\mathbb{\omega}}$ then for
some $m\in\mathbb{N}^{+}$, all $n>m$, $t$ is an element of $T_{n}$,
and all $T_{n}$ are event-algebras. That $(X_{\mathbb{\omega}},T_{\mathbb{\omega}},P_{\omega})$
is a gps follows similarly.

It remains to show that $T_{\mathbb{\omega}}$ is simple. Take ``$\neg_{n}$''
to be ``$\neg$'' defined on $T_{n}$ and ``$\neg_{\mathbb{\omega}}$''
to be ``$\neg$'' defined on $T_{\mathbb{\omega}}$. For $t\in T_{\mathbb{\omega}}$,
if $t^{\prime}\in\neg_{\omega}\neg_{\omega}\neg_{\omega}t$ then for
some $t_{1},t_{2}\in T_{\mathbb{\omega}}$, $t^{\prime}\in\neg_{\omega}t_{2}$,
$t_{2}\in\neg_{\omega}t_{1}$, and $t_{1}\in\neg_{\omega}t$, so for
some $m,n,p\in\mathbb{N}$, $t^{\prime}\in\neg_{m}t_{2}$, $t_{2}\in\neg_{n}t_{1}$,
and $t_{1}\in\neg_{n}t$, in which case, with $q=Max(\{m,n,p\})$,
$t^{\prime}\in\neg_{q+1}t$, and so $t^{\prime}\in\neg_{\omega}t$.
\end{proof}
$X_{\mathbb{\omega}}$, $T_{\mathbb{\omega}}$, and $P_{\omega}$
have clear interpretations: Given that $(X,T)$ is an event-algebra,
$X_{\mathbb{\omega}}$ is the set of of partitions whose total probability
will be $1$ for any gps on $(X,T)$. Given $X_{\omega}$, $T_{\mathbb{\omega}}$
is smallest set s.t. $(X_{\mathbb{\omega}},T_{\mathbb{\omega}})$
is an event-algebra that contains $(X,T)$. $(X_{\mathbb{\omega}},T_{\mathbb{\omega}})$
is also the largest event algebra onto which any probability function
on $(X,T)$ can be uniquely extended. Given a gps on $(X,T)$, $(X,T,P)$,
$P_{\omega}$ is $P$'s unique extension to $T_{\mathbb{\omega}}$. 

While $(X,T,P)$ and $(X_{\mathbb{\omega}},T_{\mathbb{\omega}},P_{\omega})$
are in many ways equivalent, $(X_{\mathbb{\omega}},T_{\mathbb{\omega}},P_{\omega})$
has a clearer structure, and often contains more readily accessible
information, though this is achieved at the price of sacrificing knowledge
of admissible experiments. If, for some system, you know the rules
that govern the system's probabilities, but do not know its allowed
experiments, you are likely to form $(X_{\mathbb{\omega}},T_{\mathbb{\omega}},P_{\omega})$
as its probability space, rather than $(X,T,P)$. For example, if
one were to try to deduce the allowed experiments in quantum mechanics
by analyzing quantum probability, one would arrive at $(X_{\mathbb{\omega}},T_{\mathbb{\omega}},P_{\omega})$
rather than $(X,T,P)$.

\subsection{\label{subsec:Convergence}Convergence}

When performing probability calculations, countable additivity is
generally preferred over finite additivity. Countable additivity allows
for the introduction of integration, which in turn allows probabilities
to be calculated by integrating over probability density functions.
If $(X,T,P)$ is a gps, countable additivity is the requirement that:

1) If $A\subset T$ is countable, pairwise disjoint, and $\bigcup A\in T$,
then $P(\bigcup A)=\sum_{t\in A}P(t)$.

We generally also require a guarantee as to when $\bigcup A\in T$:

2) If $A\subset T$ is countable, and for some $\gamma\in X$, $\bigcup A\subset\gamma$,
then $\bigcup A\in T$.

A gps that satisfies (1) and (2) is \emph{convergent}. Similarly,
an event-algebra is convergent if it satisfies (2). 

Reverting back to dps's for a moment, it's easy to see that convergent
dps's are equivalent to a consistent set of convergent eps's, where
convergent eps's obey the 5 original axioms listed in section \ref{subsec:Dynamic-Probability-Spaces},
but with axioms 3 \& 5 amended to hold for countable sets.

We will now show that if gps $(X,T,P)$ is convergent, then so are
$(X_{1},T_{1},P_{1})$ and $(X_{\mathbb{N}},T_{\mathbb{N}},P_{\mathbb{N}})$.
Thus, when convergence is included, the theory remains essentially
unchanged.
\begin{lem}
\label{lem: Convegence lemma}If $t_{1}\in T_{1}$ and $t\in T$ then
$t_{1}\bigcap t\in T$
\end{lem}

\begin{proof}
For some $t_{2},t_{3}\in T$, $t_{1}=t_{2}\bigcup t_{3}$. $t_{2}\bigcap t,t_{3}\bigcap t\in T$
and $t_{2}\bigcap t,t_{3}\bigcap t\subset t$ so $(t_{2}\bigcap t)\bigcup(t_{3}\bigcap t)\in T$.
Since $t_{1}\bigcap t=(t_{2}\bigcap t)\bigcup(t_{3}\bigcap t)$, $t_{1}\bigcap t\in T$.
\end{proof}
\begin{thm}
If $(X,T,P)$ is a convergent gps then $(X_{1},T_{1},P_{1})$ is a
convergent gps
\end{thm}

\begin{proof}
A: $(X_{1},T_{1})$ is a convergent event-algebra

- Take any countable $A\subset T_{1}$ and $t_{1}\in T_{1}$ s.t.
$\bigcup A\subset t_{1}$. For some $t_{2},t_{3}\in T$, $t_{1}=t_{2}\bigcup t_{3}$.
By the above lemma, and the fact that $(X,T,P)$ is convergent, $t_{4}\equiv\bigcup_{t^{\prime}\in A}t_{1}\bigcap t_{2}\in T$
and $t_{5}\equiv\bigcup_{t^{\prime}\in A}t_{1}\bigcap t_{3}\in T$.
Therefore $t_{4},t_{5}\in T_{1}$. Since $t_{4},t_{5}\subset t_{1}$,
$t_{4}\bigcup t_{5}\in T_{1}$. $\bigcup A=t_{4}\bigcup t_{5}$, so
$\bigcup A\in T_{1}$. -

B: $P_{1}$ is a countable additive

- Take any countable, pairwise disjoint $A\subset T_{1}$ s.t. $\bigcup A\in T_{1}$.
Because countable sums are defined as the limit of their partial sums,
$\sum_{t\in A}P(t)\leq1$, so $\sum_{t\in A}P(t)$ is absolutely convergent.

As $\bigcup A\in T_{1}$, take $t_{1},t_{2}\in T$ s.t. $\bigcup A=t_{1}\bigcup t_{2}$
and $t_{1}\bigcap t_{2}=\emptyset$. Because $(X_{1},T_{1}.P_{1})$
is finitely additive, $\sum_{t\in A}P_{1}(t)=\sum_{t\in A}(P(t\bigcap t_{1})+P(t\bigcap t_{2}))$.
Since $\sum_{t\in A}P(t)=\sum_{t\in A}(P(t\bigcap t_{1})+P(t\bigcap t_{2}))$
is absolutely convergent, $\sum_{t\in A}P(t)=\sum_{t\in A}P(t\bigcap t_{1})+\sum_{t\in A}P(t\bigcap t_{2})$.
Because $(X,T,P)$ is a convergent gps, $\sum_{t\in A}P(t\bigcap t_{1})=P(t_{1})$
and $\sum_{t\in A}P(t\bigcap t_{2})=P(t_{2})$. Therefore $\sum_{t\in A}P_{1}(t)=P(t_{1})+P(t_{2})=P_{1}(\bigcup A)$.
-
\end{proof}
\begin{cor}
If $(X,T,P)$ is a convergent gps then for all $n\in\mathbb{N}^{+}$
$(X_{n},T_{n},P_{n})$ is a convergent gps.
\end{cor}

We may now prove the desired result:
\begin{thm}
If $(X,T,P)$ is a convergent gps then $(X_{\mathbb{\omega}},T_{\mathbb{\omega}},P_{\omega})$
is a simple convergent gps.
\end{thm}

\begin{proof}
A: If $t\in T_{\mathbb{\omega}}$, $t^{\prime}\in T_{n}$ and $t\subset t^{\prime}$
then $t\in T_{n}$.

- Follows from Lemma \ref{lem: Convegence lemma} and simple induction
-

B: $(X_{\omega},T_{\omega})$ is a convergent event-algebra

- Take any countable $A\subset T_{\mathbb{\omega}}$, $t\in T_{\mathbb{\omega}}$
s.t. $\bigcup A\subset t$. For some $n\in\mathbb{N}^{+}$, $t\in T_{n}$.
It follows from (A) that for all $t^{\prime}\in A$, $t^{\prime}\in T_{n}$.
Since $(X_{n},T_{n})$ is a convergent event-algebra, $\bigcup A\in T_{n}$

C: $P_{\omega}$ is a countable additive

- Take any countable, pairwise disjoint $A\subset T_{\mathbb{\omega}}$
s.t. $\bigcup A\in T_{\mathbb{\omega}}$. For some $n\in\mathbb{N}^{+}$,
$\bigcup A\in T_{n}$. It follows from (A) for all $t\in A$, $t\in T_{n}$.
Because $P_{n}$ is convergent $P_{\omega}(\bigcup A)=P_{n}(\bigcup A)=\sum_{t\in A}P_{n}(t)=\sum_{t\in A}P_{\mathbb{N}}(t)$.
-
\end{proof}

\subsection{Concluding note}

The ability to run experiments and assign probabilities to their outcomes
should enable the generation and testing of scientific hypotheses.
Therefore, any world that satisfies these requirements for experiment
\& their probabilities ought to be, at least to some extent, scientifically
comprehensible.


\begin{thebibliography}{10}
\bibitem{Shoenfield}Joseph R. Shoenfield (2001) {[}1967{]}. \emph{Mathematical
Logic (2nd ed.)}, A K Peters. ISBN 978-1-56881-135-2.

\bibitem{Everett}Hugh Everett (1957). \emph{Relative State Formulation
of Quantum Mechanics}, Reviews of Modern Physics 29: 454-462.

\bibitem{Bohm 1}Bohm, David (1952). \emph{A Suggested Interpretation
of the Quantum Theory in Terms of \textquotedbl Hidden Variables\textquotedbl{}
I}, Physical Review. 85 (2): 166--179.

\bibitem{Bohm 2}Bohm, David (1952).\emph{ A Suggested Interpretation
of the Quantum Theory in Terms of \textquotedbl Hidden Variables\textquotedbl{}
II}, Physical Review. 85 (2): 180--193.

\bibitem{Durr overview}Durr, D.; Zanghi, N.; Goldstein, S. (Nov 14,
1995). \emph{Bohmian Mechanics as the Foundation of Quantum Mechanics},
\href{https://arxiv.org/abs/quant-ph/9511016}{arXiv:quant-ph/9511016}

\bibitem{Fenyes}Fényes, I. (1946). \emph{A Deduction of Schrödinger
Equation}, Acta Bolyaiana. 1 (5): ch. 2.

\bibitem{Nelson 1}Nelson, Edward (1966). \emph{Derivation of the
Schrödinger Equation from Newtonian Mechanics}, Physical Review. 150
(4): 1079--1085.

\bibitem{Nelson 2}Edward Nelson (2012). \emph{Review of Stochastic
Dynamics}, J. Phys.: Conf. Ser. 361 01201.

\bibitem{Madelung}E. Madelung (1926), \emph{Eine anschauliche Deutung
der Gleichung von Schrödinger}, Naturwissenschaften 14: 1004-1004

\bibitem{Bell 1}Bell, John S. (1982). \emph{On the Impossible Pilot
Wave}, Foundations of Physics, 12(10): 989--999. Reprinted in Bell
1987c: 159--168.

\bibitem{Shucker}Shucker, D. (1980). \emph{Stochastic Mechanics of
Systems with Zero Potential}, J. Functional Analysis 38: 146-155.

\bibitem{Ghirard}G. C. Ghirard, C. Omero, A. Rimini, I. Weber (1978).
\emph{The stochastic interpretation of quantum mechanics: A critical
review}, Riv. Nuovo Cim. 1: 1--34.

\bibitem{Bell 2}Bell, J.S. (1964). \emph{On the Einstein-Podolsky-Rosen
paradox}, Physics, 1: 195--200; reprinted in Bell 1987b {[}2004{]}:
14--21.

\bibitem{Kochen}S. Kochen; E. P. Specker (1967). \emph{The problem
of hidden variables in quantum mechanics}, Journal of Mathematics
and Mechanics, 17 (1): 59--87.

\bibitem{Norsen}Travis Norsen (2014). \emph{The Pilot-Wave Perspective
on Spin}, American Journal of Physics 82: 337-348.

\bibitem{PBR}Pusey, Matthew F.; Barrett, Jonathan; Rudolph, Terry
(2011). \emph{The quantum state cannot be interpreted statistically},
\href{https://arxiv.org/abs/1111.3328}{arXiv:1111.3328 [quant-ph]}

\bibitem{Wiseman}Antoine Tilley; Howard Wiseman (2021). \emph{Non-Markovian
wave-function collapse models are Bohmian-like theories in disguise},
\href{https://arxiv.org/abs/2105.06115}{arXiv:2105.06115 [quant-ph]}

\bibitem{GRW}Ghirardi, G.C.; Rimini, A.; Weber, T. (1986). \emph{Unified
dynamics for microscopic and macroscopic systems}, Physical Review
D. 34 (2): 470--491.

\bibitem{CSL}Pearle, Philip (1989). \emph{Combining stochastic dynamical
state-vector reduction with spontaneous localization}, Physical Review
A. 39 (5): 2277--2289.

\bibitem{Durr 2}Detlef Dürr , Sheldon Goldstein, and Nino Zanghì
(1992), \emph{Quantum Equilibrium and the Origin of Absolute Uncertainty},
Journal of Statistical Physics, 67(5): 843--907.
\end{thebibliography}
\end{document}